\documentclass[conference]{IEEEtran}

\usepackage{tikz}
\usetikzlibrary{patterns, positioning, shapes, calc}
\usepackage{amsmath}
\usepackage{graphicx}
\usepackage{listings}
\usepackage{xcolor}
\usepackage{hyperref}

\definecolor{codegreen}{rgb}{0,0.6,0}
\definecolor{codegray}{rgb}{0.5,0.5,0.5}
\definecolor{codepurple}{rgb}{0.58,0,0.82}
\definecolor{backcolour}{rgb}{0.95,0.95,0.92}

\lstdefinestyle{mystyle}{
    backgroundcolor=\color{backcolour},   
    commentstyle=\color{codegreen},
    keywordstyle=\color{magenta},
    numberstyle=\tiny\color{codegray},
    stringstyle=\color{codepurple},
    basicstyle=\ttfamily\footnotesize,
    breakatwhitespace=false,
    breaklines=true,                 
    captionpos=b,
    keepspaces=true,                 
    numbers=left,                    
    numbersep=5pt,                  
    showspaces=false,                
    showstringspaces=false,
    showtabs=false,
    tabsize=1
}
\usepackage[
  backend=biber,
  maxbibnames=99,
  url=true, doi=false,isbn=false,
  style=ieee
]{biblatex}
\AtEveryBibitem{
    \clearfield{urlyear}
    \clearfield{urlmonth}
}
\usepackage{xspace}
\usepackage{caption}
\usepackage{multirow}
\newcommand{\Code}[1]{{\texttt{#1}}\xspace}
\newcommand{\syscall}{\Code{syscall}}
\newcommand{\wrpkru}{\Code{WRPKRU}}
\newcommand{\AD}{\Code{AD}}
\newcommand{\WD}{\Code{WD}}
\newcommand{\PKRU}{\Code{PKRU}}

\newcommand{\tls}[1]{\Code{\%gs:(#1)}}
\newcommand{\Tls}{\Code{\%gs}}

\newcommand{\vsud}{\Code{nex-sud}}
\newcommand{\vrnd}{\Code{nex-rand}}
    \makeatletter
    \newcommand{\linebreakand}{%
      \end{@IEEEauthorhalign}
      \hfill\mbox{}\par
      \mbox{}\hfill\begin{@IEEEauthorhalign}
    }
    \makeatother
\author{\IEEEauthorblockN{Fangfei Yang}
\textit{Rice University}
\and
\IEEEauthorblockN{Anjo Vahldiek-Oberwagner}
\textit{Intel Labs}
\and
\IEEEauthorblockN{Chia-Che Tsai}
\textit{Texas A\&M University}
\linebreakand
\IEEEauthorblockN{Kelly Kaoudis}
\textit{High Desert Software Ltd}
\and
\IEEEauthorblockN{Nathan Dautenhahn}
\textit{Rice University/Riverside Research}
}
\date{}

\title{\Large \bf Making \`{}syscall\'{} a Privilege not a Right}

\begin{document}

\maketitle

\begin{abstract}
    Browsers, Library OSes, and system emulators rely on sandboxes and in-process isolation to emulate system resources and securely isolate untrusted components.
All access to system resources like system calls (syscall) need to be securely mediated by the application.
Otherwise system calls may allow untrusted components to evade the emulator or sandbox monitor, and hence, escape and attack the entire application or system.
Existing approaches, such as ptrace, require additional context switches between kernel and userspace, which introduce high performance overhead.
And, seccomp-bpf supports only limited policies, which restricts its functionality, or it still requires ptrace to provide assistance.

In this paper, we present nexpoline, a secure syscall interception mechanism combining Memory Protection Keys (MPK)~\cite{intel_vol3} and Seccomp or Syscall User Dispatch (SUD)~\cite{SUD:online}.
%
%
%
%
Our approach transforms an application's \syscall instruction into a privilege reserved for the trusted monitor within the address space, allowing flexible user defined policy. 
To execute a syscall, the application must switch contexts via nexpoline. 
It offers better efficiency than secure interception techniques like ptrace, as nexpoline can intercept syscalls through binary rewriting securely. 
Consequently, nexpoline ensures the safety, flexibility and efficiency for syscall interception. 
Notably, it operates without kernel modifications, making it viable on current Linux systems without needing root privileges.
Our benchmarks demonstrate improved performance over ptrace in interception overhead while achieving the same security guarantees. 
When compared to similarly performing firejail, nexpoline supports more complex policies and enables the possibility to emulate system resources.

\end{abstract}

\section{Introduction}
System call interception has been widely utilized in various computing domains, ranging from improving system safety and performance to aiding in emulation and debugging. 
Its widespread use in technologies like containers~\cite{Improvin37:online, Seccomps33:online, k8sseccomp:online, Lei2017SPEAKERSE, mine, confine} and sandbox~\cite{seccomp-sandbox-chrome:online, mbox, firejail, sysfilter, temporal}, which utilize system call interception for security purposes~\cite{Dune, sysfilter, temporal}, and in tools such as strace, which are indispensable for debugging and security modeling~\cite{Lei2017SPEAKERSE, mine, sysfilter, automodel}. 
System call interception provides security-related applications with protection against jailbreaking through the operating system and allows for complex policies.
Furthermore, its application in providing operating systems interface for libOS~\cite{Graphenelibos, Unikernels, Unikraft}, emulation purposes~\cite{WineHQRu68:online, darlingh95:online} or compatibility. 
%

\begin{figure}
    \centering
    \tikzset{cross/.style={cross out, draw, 
         minimum size=2*(#1-\pgflinewidth), 
         inner sep=0pt, outer sep=0pt}}
    \begin{tikzpicture}
            \node at (-0.7,0.5em) {User};
            \node at (-0.7,-1em) {Kernel};
            \draw[dashed] (-0.25,-0.15) -- +(-1,0);

            \draw[<-] (-0.25,-0.8) -- +(-0.8,0) node[below, midway] {syscall};
        \begin{scope}
            \draw (0,0) rectangle +(0.6,1);
            \node at (0.3,0.8) {\scriptsize app};
            \draw (0.7,0) rectangle +(0.6,1);
            \node at (0.3+0.7,0.8) {\scriptsize mon};
            \draw (0,-0.3) rectangle +(1.3, -0.7);
            \draw[dashed] (0,-0.15) -- +(1.3,0);
            \node at (0.6, -1.2) {ptrace};

            \draw[->] (0.4, 0.5) -- (0.4, -0.6);
            \draw[->] (0.5, -0.6) -- (0.9, -0.6) -- (0.9, 0.5);
            \draw[->] (1, 0.5) -- (1.1, 0.5) -- (1.1,-0.8)-- (0.2, -0.8) -- (0.2, 0.5);
        \end{scope}

        \begin{scope}[xshift = 1.8cm]
            \node at (0.65/2,0.8) {\scriptsize app};
            \draw (0,0) rectangle +(0.65,1);
            \node at (0.8+0.65/2,0.8) {\scriptsize mon};
            \draw (0.8,0) rectangle +(0.65,1);
            \draw (0,-0.3) rectangle +(1.45, -0.7);
            \draw[dashed] (0,-0.15) -- +(1.3,0);
            \node at (0.6, -1.2) {seccomp-bpf};
            
            \node[rectangle, draw] at (0.6, -0.5) {{\scriptsize BPF}}; 
            
            \draw[->] (0.45, 0.5) -- (0.45, -0.3);
            \draw[->] (0.3, -0.3) -- (0.3, 0.5);
            
            \draw[->] (0.9, -0.3) -- (0.9, 0.5);
            \draw[->] (1, 0.5) -- (1.2, 0.5) -- (1.2,-0.9) -- (0.1,-0.9) -- (0.1, 0.5);
            
        \end{scope}

        
        \begin{scope}[xshift=3.6cm]
            \node at (1.4/2,0.8) {\scriptsize app};
            \draw (0,0) rectangle +(1.4,1);
            \draw (0,-0.3) rectangle +(1.4, -0.7);
            \draw[dashed] (0,-0.15) -- +(1.4,0);
            \node at (0.7, -1.2) {nexpoline};
            \node[rectangle, draw, minimum width=1.2cm] at (0.7, -0.5) {{\scriptsize SUD}}; 

            
            \draw[fill=green!20, dashed, draw=red, thick] (1.4, 0.5) rectangle +(-0.6,-0.5);
            \draw[->] (0.5, 0.5) -- (0.5, -0.3) node[cross=2pt,red]{};
            \node at (1.4-0.3,0.3) {\scriptsize mon};

            \draw[->] (0.95, 0.7) -- (0.95, 0.5);
            \draw[->] (0.95, 0) -- (0.95, -0.3);
            
            \draw[->] (1.15, -0.3) -- (1.15, 0);
            \draw[->] (1.15, 0.5) -- (1.15, 0.7);
        \end{scope}
    \end{tikzpicture}
    \caption{Secure Syscall Interception Mechanism}
    \label{fig:system}
\end{figure}
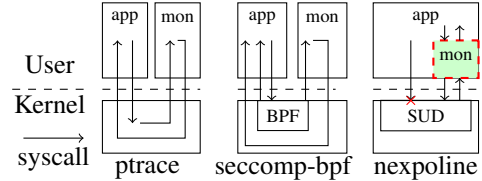

Unfortunately, existing approaches to system call interception often suffer from drawbacks related to security, efficiency, and policy flexibility.
Binary-rewrite techniques, developed through tools like syscall\_intercept~\cite{pmemsysc0:online}, e9patch~\cite{e9patch}, and zpoline~\cite{zpoline}, provided efficient redirection of programs but could be easily bypassed by jumping to existing \Code{syscall} instructions, compromising their security effectiveness.
To securely intercept syscalls, methods such as ptrace or seccomp-bpf must be used, as shown in Fig~\ref{fig:system}.
Seccomp-bpf~\cite{Seccomp} is introduced for security-oriented purposes, but it only allows BPF policy program to access only registers without the ability to preserve context information. 
Ptrace, when used alone, or in conjunction with seccomp-bpf to offer improved efficiency, but will redirected intercepted system calls to another process, leading to slower performance~\cite{gVisorCost}.

eBPF and kprobe require superuser permissions and operate globally. 
eBPF communicates with the monitor process via shared memory. 
While ideal for applications like instrumentation, it lacked the flexibility needed for complex scenarios (e.g., check file path).
And, kprobe and custom Linux Security Module~\cite{LSM} (LSM) can offer flexible policies, they need to be compiled as kernel modules and loaded, which limits their use in unprivileged applications.

Syscall User Dispatch~\cite{SUD:online} (SUD) provided a method to limit valid syscall locations, converting all other system calls into signals. 
However, it was not originally designed for security, allowing its switch bit to be flipped easily, and a jump to the legal \syscall can allow bypass of the interception.
Jenny and Enclosure, have attempted kernel modifications~\cite{Dune, Jenny, Ghosn:Enclosure:2021} for securing \Code{syscall} interception for the in-process monitor.
However, these modifications were tailored for specific scenarios rather than for a general use, posing challenges for their inclusion in the kernel~\cite{RFCPATCH40:online}.
Some modifications still necessitated a switch to the kernel mode and then a switch back~\cite{Jenny}, adding overhead from the context switches.


Our solution, nexpoline, is a trampoline mechanism that ensures efficient and non-bypassable syscall interception. 
Nexpoline divides the application into two parts: a trusted part that performs syscall interception and an untrusted part consisting of the original application. 
Based on this division, we require that \Code{syscall} instructions can only be executed by the trusted part, which includes a path that has switched through nexpoline.
In the untrusted part, syscall instructions are redirected to nexpoline for system call processing. 
This can be achieved by modifying libc or through binary rewriting.
The path integrity of nexpoline ensures that applications cannot exploit existing \Code{syscall} instructions through jumping. 
This is achieved by combining the features of MPK with Seccomp-bpf or SUD to protect and verify the memory containing the path indicator.

We explored various methods to implement nexpoline. 
While SUD requires newer kernel versions, seccomp-bpf relies on randomization for security but can be used on any kernel supporting MPK.
All these implementations require no modifications to the kernel and do not rely on global control flow integrity of main applications or ptrace and signals, which cause high performance overhead, while still offering security gains.
Our results show that, compared to other methods, nexpoline achieves secure interception of system calls with overhead similar to binary rewriting, while also offering flexible policy implementation.

\paragraph{Contribution}
In our work, we introduce a novel method to make invoking \syscall instructions a privilege in user space, ensuring secure interception of all \syscall-s from users. Key contributions of our approach include:

\begin{itemize}
    \item We identify problems in current syscall interception methods, as they fail to simultaneously achieve efficiency, security, and policy flexibility.
    \item We implemented nexpoline based on MPK, Seccomp, and SUD. It allows for secure, efficient, and flexible \syscall interception, achieved by transforming \syscall into a privilege.
    \item We implemented secure signal delivery in nexpoline, notably without requiring any modifications to the kernel.
    \item We evaluated nexpoline in both micro and macro benchmarks. We evaluated the performance of nexpoline as a sandbox for nginx and redis. Nexpoline offers better performance than traditional sandbox like mbox, while provides more flexible interception policies than firejail. 
\end{itemize}

\section{Motivation}

\paragraph{The Need for Secure System Call Interception}
In a range of application scenarios, there is a notable need for system call interception that is protected by privileged system call instructions. 
Privileged system call instructions mean that system calls can no longer be performed in any context; instead, they are restricted to a specific, authorized context.
This is particularly relevant in cases such as in-process monitors, secure environments and sandboxes like mbox and gVisor~\cite{gVisor:online}, where the system call must be restricted to protect the integrity of the monitor or the sandbox. 
These situations highlight the inadequacy of simpler methods like binary rewriting, which are susceptible to being bypassed.
To address this, technologies such as ptrace and seccomp-bpf~\cite{Seccomp} are employed~\cite{seccomp-sandbox-chrome:online, mbox}. 
They play a crucial role~\cite{chromebug:online} in restricting system calls and assigning the decision-making process to a separate, more secure mechanism. However, the implementation of these technologies can lead to additional performance overhead.
Moreover, various applications that utilize system call interception aim to prevent the bypassing of this protective layer, particularly for maintaining accurate coverage in instrumentation-focused scenarios. This concern leads some systems to adopt solutions like eBPF. 
However, eBPF's requirement for privileged loading and its global scope can be inconvenient for applications needing local interception.
Additionally, applications like Wine~\cite{WineHQRu68:online, qemu}, while not predominantly security-oriented, demonstrate the benefits of restricting native system calls to reduce their potential attack surface. 
While this approach doesn't ensure complete security, it significantly contributes to creating a more secure and manageable computing environment.

\paragraph{Applications}
With nexpoline, sandbox applications can make more decisions within the same process space. 
This allows sandboxes like mbox~\cite{mbox} to effectively intercept system calls, and to make secure decisions or implement more complex policies without fallback to ptrace.
Linux offers many features to aid in sandbox implementation~\cite{LinuxMAC}, such as LSM framework~\cite{LSM},  AppArmor~\cite{AppArmorP, AppArmor}, namespaces~\cite{namespace}, cgroups~\cite{cgroups}, SELinux~\cite{selinux}, and Landlock~\cite{landlock}. 
In fact, if the required policies can be implemented through them, the kernel can provide better performance (although not always) and assurances~\cite{LSMOverhead}, as adopted by modern sandbox implementations like firejail~\cite{firejail}. 
However, these interfaces cannot offer the same flexibility as interception. 
They only allow developers to implement policies in a specific way. 
A simple example is that you cannot use them to implement an application kernel like gVisor~\cite{gVisor:online} and support customized file systems through 9p.

In-process isolation can also benefit from nexpoline. 
For example, current works using MPK for in-process isolation require modifications to the kernel to perform system call interception for multi-threaded or signal supported applications~\cite{Jenny, ShallNot, PKRU-safe}. 
Using nexpoline can eliminate such need, thereby lowering the barrier for using in-process isolation in practice.

\section{Threat Model}
Our threat model divides the user space into trusted interceptor program and untrusted user application. 
The user application contains all user data and code, and can make invoke \Code{syscall} instructions during execution. 
Attackers have full control over the untrusted user application, which means they can perform arbitrary reads, writes, or jumps, attempting to alter the running state of the trusted program or exploit existing \Code{syscall} instructions in both trusted and untrusted's code.

We employ the same binary scanning as ERIM~\cite{ERIM} and  to prevent the presence of \Code{WRPKRU} and \Code{XRSTOR} instructions, except for in the nexpoline trampoline. Compared to the 2-byte \Code{syscall} instruction, the length of these instructions makes them easier to filter out using binary scanning.
Syscall instructions that do not pass through nexpoline will be intercepted by SUD or seccomp, even though they can be executed.
We assume that the implementation of the nexpoline is bugfree as well as the user-defined interception programs or in-process monitors. 
We assume that privileged components such as the operating system, hypervisor and hardware are trustworthy, and that nexpoline is initially loaded as the application loader. 

Side-channel attacks, Rowhammer and Denial of Service (DoS) attacks are out of our scope.

\section{Nexpoline}
\subsection{Background}
\paragraph{Syscall User Dispatch} In a nutshell, SUD is a mechanism recently added to the Linux kernel that allows for the efficient emulation of system calls within only a part of their process~\cite{SUD:online}. 
By specifying an offset, length, and selector, SUD permits system calls only from the $[\text{offset}, \text{offset}+\text{length})$ range when the value of \Code{*selector} is set to \Code{block}. 
Any system calls outside this range trigger a \Code{SYSSIG} signal. 
When \Code{*selector} is set to \Code{allow}, SUD temporarily removes the restriction and allows all \syscall, a change that can be made without a system call.
However, SUD is not a security mechanism in itself, as it only limits the IP of \Code{syscall} but cannot prevent attackers from jumping to this address or modifying the \Code{*selector} value.

\paragraph{Seccomp-bpf}
Seccomp-bpf is a security facility in the Linux kernel that allows users to load a BPF program to inspect the parameters of system calls before they are executed.
However, this BPF program has limited access to resources; for example, it can only access the values of registers used by the system call, not the memory pointed to by these registers. 
For instance, seccomp-bpf cannot filter \Code{open} based on the file's path.
For complex policies, the BPF program can only return RET\_TRAP or RET\_TRACE to handle them through signals or a ptrace debugger.
Besides accessing \syscall arguments, the BPF program can also access the instruction\_pointer, which indicates the \Code{IP} of the syscall instruction. 
Therefore, seccomp-bpf can offer policies similar to SUD, with less flexibility, such as dynamically enabled or disabled through a selector.

\paragraph{Memory Protection Keys}
MPK allows the switching of the current memory protection state through the \wrpkru instruction in the user space. 
It operates by marking each memory page with a pkey ranging from $0$ to $15$. Additionally, the CPU contains a 32-bit \PKRU register. 
For each pkey, \PKRU has two bits designated for access (data read) denial (\AD) and write denial (\WD). 
During memory access, MPK checks whether the \AD and \WD bits for the page's pkey in the \PKRU are set, in addition to the usual page protection, to determine if the operation is allowed.
In nexpoline, we use 3 pkeys, 0 and 1 for the memory used by nexpoline and the 2 for user. 
Their allocation is shown in Table~\ref{tab:mpkey}.
\begin{table}[t]
    \centering
    \caption{MPK Permission Assignment}
    \begin{tabular}{c|c|c|c}
        \multirow{2}{*}{PKRU} & \multicolumn{3}{c}{pkey} \\
         & 0 & 1 & 2 \\ \hline
        \Code{\$trusted\_pkru} & rw & rw & rw \\
        \Code{\$untrusted\_pkru} & ro & - & rw \\
    \end{tabular}
    \label{tab:mpkey}
\end{table}
\Code{\$trusted\_pkru} refers to a PKRU that allows access and write to both memory. 
\Code{\$untrusted\_pkru} refers to a PKRU that permits access and write to user memory, access to nexpoline memory with pkey 0 and no access with pkey 1.

In Linux, a new pkey can be created using \Code{pkey\_alloc}, and the pkey of a page can be changed with \Code{mprotect\_pkey}.

\subsection{Design}
\newcommand{\boldCode}[1]{\textcolor{magenta}{\fontfamily{lmtt}\fontseries{b}\selectfont #1}}
\begin{figure}
    \centering
    \begin{tikzpicture}[remember picture]
        \node[rectangle, rounded corners, align=left, draw=black, minimum width=3cm, minimum height=1.4cm] (Source) at (0,0) { \tikz[remember picture]\coordinate(scL){};\boldCode{syscall}\tikz[remember picture]\coordinate(scR){}; \\ \tikz[remember picture]\coordinate(CallS){};\boldCode{call }\Code{trap}};
        \draw [red] ([yshift=0.25em, xshift=-0.2em]scL) -- ([yshift=0.25em, xshift=0.2em]scR);

        \node[rectangle, rounded corners, align=left, draw=black, minimum width=3cm, text width=3cm, minimum height=3cm, below= 2em of Source.south]  (Nexpoline) {\tikz[remember picture]\coordinate(TrapS){};\Code{trap:}\\\;\;\boldCode{mov }... \\\;\;\boldCode{WRPKRU} \\\;\;\boldCode{mov }\Code{0, *Sel}\tikz[remember picture]\coordinate(SelCodeE){};\\\;\;\boldCode{call}\Code{ Intercept}\tikz[remember picture]\coordinate(CallintE){};\\\;\;\boldCode{mov}\Code{ 1, *Sel}\\...};

        \node[very thick, rectangle, rounded corners, align=left, draw=red, right= 1.6em of Source.north east, anchor=north west, inner sep=0.6em] (Selector) {Selector = 1};

        \node[thick, rectangle, rounded corners, align=left, draw=black, fill=gray!10, text width=3.4cm, minimum width=3.4cm, minimum height=2cm,anchor=north, below=0.5em of Selector.south, xshift=1.2cm] (KernelCode) {\Code{Kernel\_Entry()}\\
        \;\;\Code{if Sel==0}\tikz[remember picture]\coordinate(SelE){};\\
        \;\;\;\; Normal Syscall\\
        \;\;\Code{else}\\
        \;\;\;\; Failed Signal
        };

        \node[thick, rectangle, rounded corners, align=left, draw=red, minimum width=3.4cm, text width = 3.3cm, minimum height=1cm,below= 1em of KernelCode ] (IntCode) {
        \Code{Intercept()}\\
        \;\;\;\;\Code{... Handling }\\
        \;\;\;\;\tikz[remember picture]\coordinate(SyscallS){};\boldCode{syscall}\\
        \;\;\;\;\Code{...}
        };

        \node[above = 0.25em of Selector, text=red] {MPK Protected};
        
        \draw[<-, red] ([xshift=0.25em]Selector.east) -- + (1.5em, 0) node [anchor=west] (Proof) {Proof};
        \draw [->, red] (Proof.south) |-  ([yshift=0.25em]SelE) node[xshift=1.2cm,yshift=-0.5em] {Verify};
        
        \node[above = 0.25em of Source] {Source Code};
        \node[above = 0.25em of Nexpoline] {Nexpoline};

        \draw[->, red] ([yshift=0.25em]SelCodeE) -| ([xshift=-1cm, yshift=-0.1em]Selector.south) node[midway, sloped, pos=0.8, above] {Set SUD Selector};

        \draw[->] ([yshift=0.25em, xshift=-0.25em]SyscallS) -| ([xshift=-0.65em]KernelCode.west) -- +(0.5em, 0);

        \draw[->] ([yshift=0.25em]CallintE) -- ([yshift=0.25em, xshift=-0.1em]CallintE-|IntCode.west);

        \draw[->] ([yshift=0.25em]CallS) to[out=190, in=90] ([yshift=0.75em, xshift=0.2em]TrapS);
    \end{tikzpicture}
    \caption{Nexpoline Overview}
    \label{fig:arch}
\end{figure}
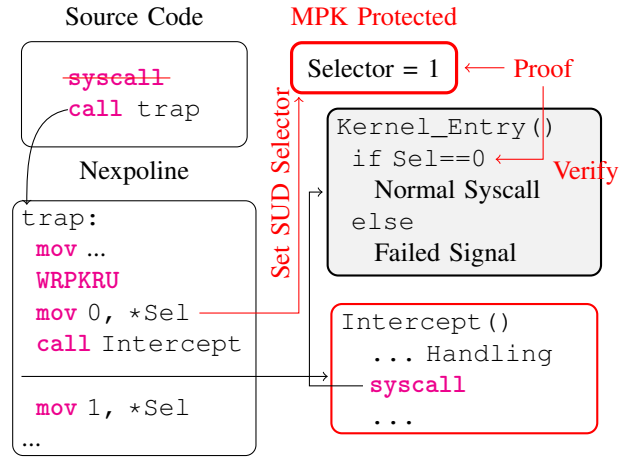

\textbf{Overview} -- Intercepting system calls within the process can reduce the additional performance overhead caused by process switching in methods such as ptrace. At the same time, unlike BPF programs which need to consider kernel security, user programs face fewer restrictions, making  complex policies feasible. 

To achieve this, nexpoline divides the process space into two parts: the untrusted user application and the trusted system call interceptor program. 
Nexpoline's trampoline acts as a callgate bridging these two, forcing the user application to switch to the trusted part through nexpoline before making system calls. 
SUD or seccomp is used to verify that the syscall originates from the trusted part.
It effectively makes invoking \Code{syscall} instruction  a privilege and ensure the interception of all system calls. 
Also, nexpoline redirects signal handlers, ensuring that the running state of the trusted part is not exposed to the user application through signals. 
Lastly, nexpoline implements system call policies to protect the integrity of MPK.

\textbf{Nexpoline} --
Nexpoline works through the combination of two features to ensure syscall instructions cannot be used by the untrusted part. 
SUD or Seccomp-bpf is used to prohibit illegal system calls. 
Fig~\ref{fig:arch} demonstrates the basic structure of nexpoline when using SUD. 
The value of \Code{*selector} determines if system calls are allowed. 
After the MPK switch in the nexpoline, we set \Code{*selector} to 0, allowing the subsequent interception program to make system calls, and restore this value to 1 afterward. 
This ensures that only programs that have passed through nexpoline can make system calls.
With Seccomp-bpf, we use a BPF program to determine whether to allow system calls, based on whether the instruction pointer is within the gadget segment, whose address is within the range of $[sbegin, send)$.
This region is mapped and protected at the start of the application's execution and is never unmapped. 
During the nexpoline switch, we randomly put a 3-byte gadget \Code{syscall;ret} within the $[sbegin, send)$ range that is not in use, and save the gadget address. 
We then use this randomized address to make syscall, and upon returning from nexpoline, the gadget is reset to \Code{int3}-s.
Unlike SUD, Seccomp-bpf cannot reset different filters for new threads. 
Therefore, all threads must use the fixed IP or IP range for filtering \syscall address. 
If a fixed address is used, one thread's nexpoline allows another thread to use the same \Code{syscall;ret} for its system call, regardless of whether it is trusted or untrusted.
Our solution allows all \Code{syscall} instructions within a gadget segment but generates a random address for each interception to prevent other threads from using it.
MPK is used to protect the memory pointed to by the selector or the gadget segment. 
For the SUD's selector, we can use pkey 0, which is read-only for the untrusted part, as we only need to ensure that the attacker cannot modify the selector. 
For the gadget segment, we use pkey 1, which the untrusted part cannot read, to prevent attackers from finding the \Code{syscall;ret} location.

\textbf{Signal Handling} --
Nexpoline needs to redirect signals, as the kernel can transfer control to the signal handler at any time, including when running in the trusted part or the nexpoline. 
When the user program is running in the untrusted part, our handler only needs to give the control back to the user's signal handler, as the kernel has already correctly pushed the sigframe onto the appropriate position in the user stack.
When the user program runs in the trusted part, which means we are processing a syscall interception, we assume that the signal occurred at the time the user requested the \syscall. 
We generate a new sigframe on the user's stack based on that. 
And, when returning from nexpoline, we return to the user's signal handler.
%
%

%
Nexpoline performs context switching in stages, which means that intermediate states can also lead to signal distribution. We need to analyze which information is trustworthy in each state in order to handle it correctly.

\textbf{Signal Return} --
Nexpoline must handle user's sigreturns, as this system call changes the control flow, affecting nexpoline's action of clearing the selector or syscall gadget upon return.
Nexpoline reads the sigframe on the user stack and copy the registers.
Subsequently, these registers are restored during the return process of nexpoline, and we ensure that PKRU is also correctly set to the user's PKRU.

\section{Implementation}

Nexpoline's implementation is challenging when considering performance, compatibility, and security. To allow for easy adoption, our implementation requires no kernel modifications. Our implementation particularly focuses on the following three key goals of Nexpoline.
\begin{enumerate}
    \item Our mechanism needs to be viable in a multi-threaded environment and ensure security.
    \item We need to ensure secure handling of signals, which includes signals related to MPK, SUD and Seccomp, as well as preventing signals from affecting the interception process.
    \item We need to prevent known MPK-related attack methods from impacting the integrity of nexpoline.
\end{enumerate}

\label{sec:impl}
\subsection{Multithraeding Support}
Nexpoline needs to consider the multi-threaded environment because each thread has its own stack, signals, and other running states.
\textbf{Threading Context Initialization} --
We create a threading context protected by MPK for each thread and place its address in \Tls. 
%
We need to do this for each new thread created through \Code{clone}.

\textbf{Thread-Independent Syscall Switch} --
SUD is enabled separately in each thread and sets a different \Code{selector} address. 
By changing \Code{*selector}, we can disable or enable system calls.
Since each thread has a different address, this prevents a system call enabled by one thread from being exploited by another. 
Seccomp-bpf does not support modifying filters for each thread, all threads use the same BPF program and hardcoded gadget segment. 
We adopt a method of syscall addresss randomization, allowing each thread to use an independent \Code{syscall;ret} address.
Before making a system call, each thread generates its own random and unique syscall gadget address, and uses this address to perform the system call.
This address quickly becomes invalid, and a different address is used for the next \syscall. 
The memory segment is unreadable (using AD from MPK), so attackers can only guess the address.
However, incorrect guesses will trigger \Code{int3}, thus being detected by nexpoline.

\subsection{Secure Signal Handling}
Signals from the kernel can interrupt the current execution state at any time, causing uncontrollable changes in the control flow. 
We need to ensure that these changes do not affect the integrity of nexpoline. 
For example, executing the user's handler in a trusted state. 
Additionally, the current version of the kernel cannot correctly deliver signals on sigaltstack. 
To avoid modifying the kernel, we need to ensure under safe conditions that the address pointed to by \Code{\%rsp} is always writable, which brings additional challenges. 
Because we will not be able to fully trust the sigframe information from the user stack.

\paragraph{Signal Aware Nexpoline} --
\begin{figure}[t]
\begin{lstlisting}[language={[x86masm]Assembler}, caption=Nexpoline Trampoline Code, label=lst:nexpoline]
nexpoline_trap: 
    // Push registers to the stack
    // Nexpoline
    // Normal MPK Switch
    xor %ecx, %ecx
    xor %edx, %edx
    mov $trusted_pkru, %eax
    wrpkru // MPK Switched
    // Stage 0: PKRU=T, Stack=U, Sel=Block
    cmp $trusted_pkru, %eax
    jne __exit
    xchg %rsp, %gs:(Scratch)
    // Stage 1: PKRU=T, Stack=T, Sel=Block
    movl $0, %gs:(Selector)
    call syscall_intercept
    movl $1, %gs:(Selector)
    xchg %rsp, %gs:(Scratch)
    // Stage 2: PKRU=T, Stack=U, Sel=Block
    // Normal MPK Switch
    xor %ecx, %ecx
    xor %edx, %edx
    mov $untrusted_pkru, %eax
    wrpkru // MPK Switched
    cmp $untrusted_pkru, %eax
    jne __exit
    // Stage 3: PKRU=U, Stack=U, Sel=Block
    // Registers restor, clear redzone
    retn 128
\end{lstlisting}
\end{figure}

Listing~\ref{lst:nexpoline} shows the simplified nexpoline trampoline code which performs MPK and context switching. 
User registers are saved on the caller's stack as a trapframe, and these registers are restored when returning from nexpoline.
We use the \wrpkru + \Code{cmp} code given in ERIM~\cite{ERIM} to ensure the correct PKRU is written, preventing attackers from jumping to the \wrpkru in nexpoline to write their own PKRU values.
To allow the signal delivery during nexpoline, we carefully arrange switches of MPK and the stack, ensuring the \Code{\%rsp} is switched only after the target stack is writable.
%
%
This means that, after switching PKRU to trusted, there will be several instructions executed with \Code{\%rsp} pointing to the user stack. 
Likewise, before switching \Code{\%rsp} back to untrusted, there are also several instructions where \Code{\%rsp} has already switched back to the user stack. 
In both scenarios, we cannot trust the registers saved sigframe on the user stack, including \Code{\%rip} and \PKRU.
Therefore, when return from the signal handler, we need to set \PKRU to the untrusted and roll back \Code{\%rip} before the \wrpkru switch. 

For signals delivered between Stage 0-Stage 1 in Listing~\ref{lst:nexpoline}, we process it as a user signal, but we can set \PKRU to trusted during \Code{sigreturn} and move \Code{\%rip} to a point after Stage 1. 
For signals delivered between Stage 1-Stage 2, they are treated as being in the trusted state because \Code{\%rsp} points to the MPK-protected stack.
For signals delivered between Stage 2-Stage 3, we can directly set PKRU to untrusted, move \Code{\%rip} to the end of Stage 3, and handle them as signals delivered in the user.
When using SUD, the trampoline switches \tls{selector} directly. 
For seccomp, the syscall gadget used by seccomp need to avoid conflicts between different threads. 
The switch requires generating a new gadget address, using atomic instructions to check and occupy the position, and finally writing the gadget to the new location.

\textbf{Secure Signal Handler} -- 
\begin{figure}[t]
\begin{lstlisting}[language={[x86masm]Assembler}, caption=Signal Entrance Trampoline Code, label=lst:sigentry]
signal_entry:
    // Kernel set PKRU=0x5555555c
    // And mask all signals
    mov 1, %gs:(Signal)
    xor %ecx, %ecx
    xor %edx, %edx
    mov $trusted_pkru, %eax
    wrpkru
    cmp $trusted_pkru, %eax
    jne __exit
    cmp 1, %gs:(Signal)
    jne __exit
    mov $0, %gs:(Selector)
    xchg %rsp, %gs:(SigStack)
    call signal_handler
    ret // -> sigreturn to interceptor
\end{lstlisting}
\end{figure}

When registering a signal, we use \Code{signal\_entry} from Listing~\ref{lst:sigentry} instead of the original sigaction function.
We set \Code{sa\_mask} to ensure that the kernel blocks re-entry into our handler function.

The code in Listing~\ref{lst:sigentry} first set \tls{Signal} before \wrpkru to avoids abuse (jump to) of this entry point by user programs. This is because the kernel automatically sets PKRU during signal delivery, ensuring that \tls{Signal} can be written to. A user program that abuses the entry point, however, cannot provide the correct PKRU needs for writing to \tls{Signal}.

Our \Code{signal\_handler} function processes signals related to MPK, and delivers other signals to the user-specified sigaction. 
At this time, there are two cases:
\begin{enumerate}
\item \Code{\%rsp} points to memory on the user stack. In this case,
we push a trapframe onto the user's stack and adjust \Tls and \Code{\%rsp}. Then, we fake a system call to sigprocmask to unmask the sigmask.  The user's signal handler will be called after the execution of the faked sigprocmask.
\item \Code{\%rsp} points to memory on a stack protected by MPK. 
In this scenario, we re-deliver the signal to the user stack, as illustrated in Fig~\ref{fig:signalrd}, which depicts the stack manipulation of signal re-delivery. 
This is achieved by pushing a new sigframe onto the user's stack, copying the contents of the current trapframe into this sigframe, and creating a new trapframe that points to this new sigframe with the real sigaction set as its \Code{\%rip}. 
Subsequently, we use sigreturn to continue the previous interception. 
The signal delivery takes place after the interception has concluded.
\end{enumerate}
Distinguishing between these is important, as it determines whether our handler function can trust the information on the signal stack, which includes the \Code{\%rip} and PKRU values where the signal occurred.

\pgfdeclarepatternformonly{north east lines sparse}{\pgfqpoint{-1pt}{-1pt}}{\pgfqpoint{10pt}{10pt}}{\pgfqpoint{9pt}{9pt}}%
{
  \pgfsetlinewidth{0.4pt}
  \pgfpathmoveto{\pgfqpoint{0pt}{0pt}}
  \pgfpathlineto{\pgfqpoint{9.1pt}{9.1pt}}
  \pgfusepath{stroke}
}

\begin{figure}[t]
    \centering
    \begin{tikzpicture}
        \draw [very thick] (0,0) -- (0,-5) -- (2.5,-5) node[midway, below] {Stack} -- (2.5,0);
        \draw [thick] (0, -4.5) -- (2.5, -4.5);
        \draw[draw=red, thick, pattern=north east lines sparse, pattern color=red] (0.1,-4.4) rectangle +(2.3,0.4);

        \draw [thick] (0.1, -3.9) rectangle +(2.3, 1.3);
        \draw [thick] (0.1, -3.5) -- +(2.3, 0);
        \node[rectangle, align=center] at (1.25, -3.7) {Old \Code{\%rip}};
        \node[rectangle, align=center] at (1.25, -3.05) {OldTrapFrame};

        \draw [thick, fill=gray!20] (0.1, -2.5) rectangle +(2.3, 0.9);
        \node[rectangle,align=center] at (1.25, -2.05) {SigFrame}; 
        \draw [fill=gray!20] (0.1, -1.5) rectangle +(2.3, 0.4);
        \draw [thick, fill=gray!20, pattern=north east lines sparse, pattern color=red] (0.1, -1.5) rectangle +(2.3, 0.4);

        \draw [thick, fill=gray!20] (0.1, -1) rectangle +(2.3, 1.3);
        \draw [thick] (0.1, -0.6) -- +(2.3, 0);
        \node[rectangle,align=center] at (1.25, -0.8) {\Code{\%rip}};
        \node[rectangle,align=center] at (1.25, -0.15) {TrapFrame};

        \begin{scope}[overlay]
            \node[very thick, rectangle, minimum width=2cm, minimum height=0.5cm, draw=black] (Scratch) at (-2,-2.6) {Scratch};
            \draw[<-] (-0.1,-2.6) -- +(-0.9,0) node[above, midway] {RSP};
            \draw[<-] (-0.1, -4.5) -- +(-0.9,0) node[anchor=east] {RSP$_{\text{USER}}$};

            \draw[<-] (-0.1, -4.2) -- +(-0.9,0) node[anchor=east] {Red-Zone} ;
        \end{scope}

        \begin{scope}[overlay, red]
            \draw [->, thick] (2.0, -0.8) -- (3.2, -0.8) node[anchor=west] {Signal Handler};

            \draw [->, thick] (0.3, -0.15) -- (-0.5, -0.15) -- (-0.5, -1.6) -- (-0.1, -1.6);

            \draw [->, thick] (-1.5,-2.2) -- (-1.5,0.3) node[above, sloped, midway] {New RSP} -- (0,0.3) ;

            \draw [->, thick] (2.3, -3.25) -- (3.3, -3.25) -- (3.3, -2.05) node[midway, sloped, below] {Copy} -- (2.2, -2.05);

            \draw [<-, thick] (2.2, -0.15) -- (3.2, 0.25) node[anchor=west] {New TrapFrame};

            \draw [<-, thick] (2.6, -1.3) -- (3.2, -1.3) node[anchor=west] {Red-Zone};
        \end{scope}
        
    \end{tikzpicture}
    \caption{User Stack Manipulation for Signal Re-Delivery during interception}
    \label{fig:signalrd}
\end{figure}
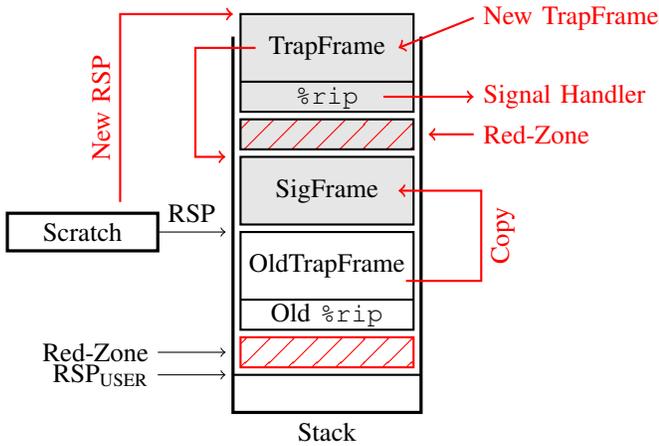

\Code{sigaltstack} could be used to avoid this distinction, thereby distributing all signals to a protected stack. 
However, it cannot be utilized due to a flaw in the current kernel implementation~\cite{PKUusage61Bug:online}.

For signal returning to the user, we cannot implement it by using a \Code{sigreturn} system call. 
%
This is because to call \Code{sigreturn}, we need to allow system calls by setting \tls{Selector} to 0. 
However, after \Code{sigreturn}, execution continues as the user, and we do not have the opportunity to reset \tls{Selector} to 1.
Therefore, we have to to emulate a \Code{sigreturn}. And, we need to copy the registers from the sigframe to the trapframe and perform \Code{xrstor} without changing the PKRU. 
Then, the correct switching of PKRU and \tls{Selector} is achieved through the normal return of nexpoline.

It's noteworthy that kernel may deliver signal during the process of copying the registers to \Code{trapframe}. 
In this case, the signal handler cannot use the registers from the \Code{trapframe} on the stack; instead, it needs to complete the copying of the registers first.

\subsection{MPK-related Attacks}
The impact of system calls on MPK security has been  studied priviously~\cite{ShallNot, Pitfalls}. 
Simply put, the kernel also adheres to MPK restrictions, but some system calls can bypass or change these limitations. 
On the other hand, other system calls might change the control flow outside of signals, such as \Code{rseq}.
Therefore, we must impose restrictions on them to ensure that nexpoline is not affected by their influence.

\begin{figure}[t]
\begin{lstlisting}[language={[x86masm]Assembler}, caption=System call delegation, label=lst:delegate]
delegated_syscall:
    mov 1, %gs:(Delegation) 
    // Provide a user stack
    xchg %rsp, %gs:(DeleStack)
    xor %ecx, %ecx
    xor %edx, %edx
    mov $untrusted_pkru, %eax
    wrpkru
    cmp $untrusted_pkru , %eax
    jne __exit
    syscall
    xor %ecx, %ecx
    xor %edx, %edx
    mov $trusted_pkru, %eax
    wrpkru
    cmp $trusted_pkru , %eax
    jne __exit
    cmp 1, %gs:(Delegation)
    jne __exit
    xchg %rsp, %gs:(DeleStack)
    mov 0, %gs:(Delegation)
    ret
\end{lstlisting}
\end{figure}

\textbf{Syscall Delegation} -- 
The nexpoline has a trusted \PKRU that allows kernel to access all memory. 
This means we need to check all pointers coming from the user to avoid unwanted access, or alternatively, we can temporarily switch \PKRU back to untrusted.
%
We refer to such system calls as delegated syscalls in Listing~\ref{lst:delegate}.

To achieve this, we need to switch PKRU back to untrusted and \Code{\%rsp} to a temporary user stack to allow signal delivery during the delegation.
\tls{Delegation} is set and check 
to prevent attackers from abusing the \wrpkru back to the trusted with a following \Code{ret} sequence in the end of Listing~\ref{lst:delegate} by jumping to it.
Signals delivered during the delegation process require special handling. 
The signal handler uses \tls{Delegation} to determine whether the signal occurred while truly in a trusted state and decides the state after sigreturn based on it. 
As we can't trust the registers in the sigframe on the temporary stack, and we need to restart the delegated syscall.
%

\textbf{Syscall Intercrption} -- In our current implementation, we rewrite the syscalls in the libraries to call to syscall\_trap.
This could be done with other binary-write based syscall interception.
In this case, we only need to change their interception entry point to nexpoline.
Regardless of the approach used, both SUD and Seccomp can prevent unauthorized syscalls, thereby enforcing the subsequent policy.

\textbf{Necessary System Call Security Policy} -- 
We need to restrict the functionality of several system calls:
a)  the target address and pkey during system calls that change page permissions, such as \Code{mprotect}, \Code{mmap}, \Code{munmap}, \Code{pkey\_alloc}, \Code{pkey\_free} etc.;
b) thread context initialization during the \Code{clone} system call;
c) rewrite the functionality of system calls that involve signals;
d)  \Code{rseq}, \Code{userfaultfd}, \Code{process\_vm\_*}, \Code{prctl}, \Code{arch\_prctl}, \Code{set\_thread\_area}, \Code{ptrace}, \Code{seccomp}, \Code{modify\_ldt}, and instructions related to shared memory. 
All these system calls may circumvent MPK protections, modify \Tls, or they might affect the integrity of the control flow.
In addition, we rely on ERIM's binary scanning to ensure that there are no \wrpkru or \Code{XRSTOR} instructions in the code during \Code{mmap} and \Code{mprotect}
and limit the use of exec by ensuring the interceptor is still loaded in the target program.

\section{Evaluation}
\subsection{Environment Setups}
Our evaluation platform uses an Intel i7-1165G7 @ 2.80GHz with Turbo Boost disabled, 16GB of memory, Ubuntu 20.10, and kernel version 5.9 (with SUD patch~\cite{patchSUD}). 
Our experiments rely on Nginx~\cite{nginx} 1.18.0 , Redis~\cite{redis} 6.2.6, ApacheBench~\cite{apachebench} 2.3, mbox~\cite{mbox} a131424 and firejail~\cite{firejail} 0.9.73 for benchmarking. 
%

%
Mbox is a classic sandbox that uses ptrace and seccomp for system call interception, while firejail employs kernel features namespaces and LSM for isolation offering the least observable performance overhead, but unfortunately failing to provide flexible enough access permission policies to support system call emulators like gVisor. We provide this baseline as a goal post to identify the least overhead nexpoline could possibly have in a sandbox application.

We have two implementations of nexpoline: one using the newer system call SUD, which we denote as \vsud, and another using Seccomp-bpf with randomized syscall gadget addresses, denoted as \vrnd.

\subsection{Security Evaluation}

In this section, we will discuss the security aspects of the different components involved in nexpoline, including the trampoline, signals, multi-threading, and the attack surfaces introduced by MPK, as well as the necessary security policies to mitigate them.

\paragraph{Nexpoline} We need to ensure that Nexpoline does not lead to the leakage of MPK privileges. The only indirect jump in our trampoline is \Code{ret}. Which means We only need to ensure that the MPK state is always untrusted before \Code{ret}, which is achieved through ERIM~\cite{ERIM}'s \wrpkru + \Code{cmp}. 
If an attacker jumps to \wrpkru with an abnormal target PKRU value, this will be captured by the following \Code{cmp} instruction and jump to an \Code{int3} instruction to terminate the process.

\paragraph{Randomization} Using randomization does not guarantee the security of \Code{syscall}. We improve this by filling the memory segment with \Code{int3} and limiting the maximum number of threads allowed to exist. 
Attackers have only one chance to find the \Code{syscall} instruction. 
The success rate is $n/size$, where $n$ is the maximum number of active threads, and $size$ is the length of the syscall segment.
Some applications use thread pools to determine the maximum number of threads, which allows us to define the necessary size of the syscall segment based on this number. Other applications may have more threads, requiring an increase in the segment size accordingly. 

\paragraph{Signal} Signals interrupt execution, and the state prior to the signal is saved on the stack pointed to by \Code{\%rsp}. Our principle is that if the memory pointed to by \Code{\%rsp} belongs to the trusted segment, we consider its content trustworthy, allowing us to return to the previous state via sigreturn. If \Code{\%rsp} points to untrusted memory, which often means the interruption occurred during the execution of user code, we simply need to return to the relevant handler. Special cases include during nexpoline switching and syscall delegation. 

In these instances, the untrusted sigframe stores the trusted PKRU, but we cannot trust the saved \Code{\%rip} value.
For nexpoline, we need to discuss the different stages of the current switch, as the stack switch occurs immediately after the MPK switch. 
We just need to ensure that both these steps are fully executed to establish a correct environment for the trusted program's execution.
The situation when returning from nexpoline is similar.
For delegation, we establish trust through \tls{Delegation}, and we do not use any register values from the sigframe.

The entry point of the signal handler sets the \tls{Signal} using defualt \PKRU from the kernel, ensures that the signal is from the kernel and not forged and abused by a user program.

\paragraph{MPK} 

\begin{table}[t]
    \centering
    \caption{Security analysis of the known MPK attakcs}
    \begin{tabular}{p{4.5cm}|p{3.5cm}}
       \multicolumn{1}{c|}{Attacks}  & \multicolumn{1}{c}{Mitigation} \\ \hline
        Change page permissions~\cite{Pitfalls} & Limited \Code{mprotect} \\ \hline
        Bypass MPK with syscall~\cite{Pitfalls} & Limited \syscall-s and syscall delegation \\\hline
        Change PKRU with Instructions ~\cite{Pitfalls} & Ban \wrpkru and \Code{xrstor} \\ \hline
        Modify Executable Page~\cite{Pitfalls, ShallNot} & No shared memory for executable-only page \\ \hline
        Utilize debug and seccomp APIs~\cite{Pitfalls} & Ban \syscall-s \\\hline
        Signal Context Attack~\cite{Pitfalls} & Trust sigcontext only when \Code{\%rsp} is safe\\ \hline
        Hijack Control-flow with Signal~\cite{Pitfalls} & Redirected signal handler \\\hline
        Change PKRU using sigreturn~\cite{Pitfalls} & Emulated sigreturn \\\hline
        Abuse Nexpoline's Signal Handler Entry & Set \tls{Signal} flags before \wrpkru \\\hline
        Other Control-flow hijacks & Ban \syscall-s\\ \hline
    \end{tabular}
    \label{tab:seceval}
\end{table}
The security of MPK is affected by the presence of instructions like \wrpkru in the code, signals and by system calls, for which we use binary scanning and related security system call policies, respectively. 
These measures can mitigate the known MPK attacks listed in Table~\ref{tab:seceval}.
\subsection{Syscall Interaction Overhead}

\begin{table}[tb]
    \centering
    \caption{CPU Cycles for Syscall Interaction}
    \begin{tabular}{c|cccccc}
        Mechanism & getpid & read & write & mmap & open & close \\ \hline
        \Code{syscall} & 215 & 289 & 242 & 801 & 2435 & 824 \\
        \Code{sec-bpf} & 281 & 366 & 303 & 868 & 2661 & 940 \\
        \Code{sud-bypass} & 297 & 375 & 324 & 917 & 2526 & 1006 \\
        \Code{sud-signal} & 5201 & 5403 & 5128 & 7583 & 12398 & 5319 \\
        \Code{ptrace} & 33072 & 34230 & 34402 & 36184 & 38103 & 35867 \\ \hline
        \Code{mbox} & 343 & 445 & 376 & 40258 & 42628 & 999 \\ 
        \Code{firejail} & 368 & 462 & 392 & 1003 & 2857 & 1077 \\ \hline
        \Code{\vsud} & 457 & 539 & 484 & 2741 & 3385 & 1081 \\
        \Code{\vrnd} & 462 & 549 & 500 & 2705 & 3582 & 1157 \\
    \end{tabular}
    \label{tab:microperf}
\end{table}

\begin{figure}[tb]
    \centering
    \hspace*{-1cm}
    \includegraphics[width=0.45\textwidth]{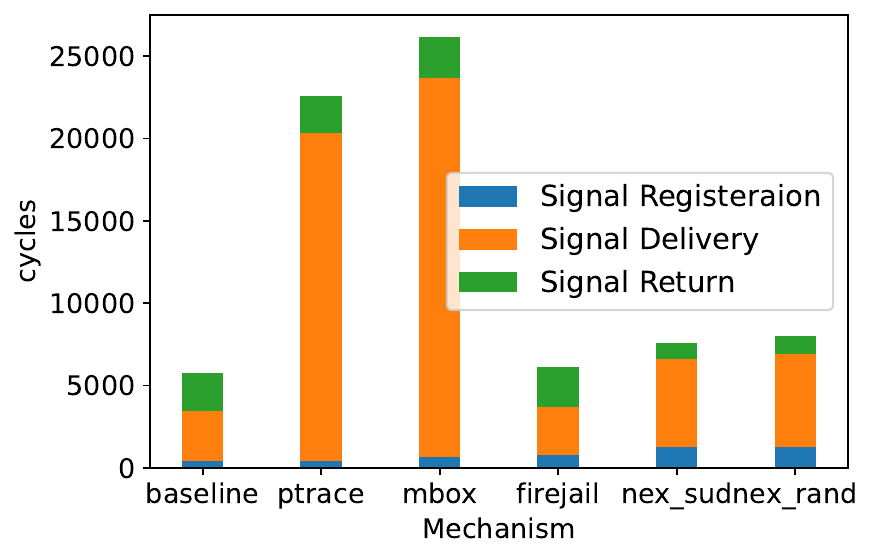}
    \captionof{figure}{Cycles on signal register, deliver and return}
    \label{fig:sigoverhead}
\end{figure}

We evaluated the overhead of secure system call interception using \vsud and \vrnd, comparing them with other syscall interception mechanism.
We use \Code{rdtsc} to measure the overhead of system calls before and after executing \Code{syscall\_trap}
Because system calls only constitute a small part of an application, their overhead tends to be obscured in the macro benchmarks. 
Such analysis helps in understanding the true overhead of the syscall interception.
We selected the syscalls in Table~\ref{tab:microperf} to cover different operating costs of nexpoline. \Code{getpid} is the fastest and the overhead of syscall interception is significant. \Code{read} and \Code{write} a 4k page on \Code{/dev/null} have medium cost, but nexpoline has no additional policy for them. \Code{mmap} and \Code{open} are slow, and nexpoline needs to perform extra checks, thus resulting in more cycles added. \Code{close} is also slow, but nexpoline does not need to check it, result in lower overhead than \Code{open}. 
%
Nexpoline makes changes in signal delivery, we also evaluated its overhead.
We trigger signal delivery through the \Code{kill} system call and evaluate the overhead of signal delivery by recording the current \Code{rdtsc} before executing the \Code{kill}, when the signal handler is invoked, and upon returning to the end of the \Code{kill}.
In Table~\ref{tab:microperf}, \Code{syscall} is the baseline. \Code{sec-bpf} shows the cycles consumed when using a seccomp policy that only allows this system calls. \Code{sud-bypass} indicates the cycles consumed when SUD is enabled but all system calls are allowed. \Code{sud-signal} denotes the consumed cycles when intercepting system calls with a signal and executing them. \Code{ptrace} represents the cycles consumed when intercepting all system calls using ptrace.

For most system calls, the context switch in nexpoline combined with the use of SUD or seccomp results in a total overhead of approximately $250$ cycles. This fixed cost is from the trampoline. 
When using syscall delegation, it consumes an additional $60$ cycles. In some cases, nexpoline requires more system calls; for example, with \Code{mmap}, nexpoline needs an additional \Code{mprotect} to set the page's pkey. For \Code{open}, nexpoline requires \Code{stat} to check if \Code{/proc/self/mem} is opened. Therefore, the overhead in these cases is higher.
Overall, nexpoline is $10x \sim 72x$ times faster than their competitions offering the same functionality and security. And only $1.0x \sim 2.7x$ times slower than the less flexible alternatives.

Fig~\ref{fig:sigoverhead} compares the cycles consumed for the signal.
Due to the re-delivery, nexpoline requires more cycles to deliver signals. Surprisingly, handling sigreturn in userspace and not relying on the kernel in nexpoline regains some of the performance reducing the overall overhead. 
As a result, for a single signal delivery, both ptrace and mbox have an overhead of $>300\%$, while \vsud and \vrnd have overhead of $17.8\%$ and $25.6\%$, respectively. Firejail has an overhead of about $\sim1\%$ which is close to seccomp only. This is because both cannot change the behavior of system calls, but only filters out illegal onces. 

\subsection{Applications Benchmarks}
In this section, we use nexpoline to implement an application sandbox, thereby isolating NGINX and Redis. 
We forbid most of the unused system calls and limit file system interactions. 
We compared our results with mbox and firejail. 

\textbf{NGINX Server} -- 
\begin{figure}[t]
    \centering
    \includegraphics[width=0.45\textwidth]{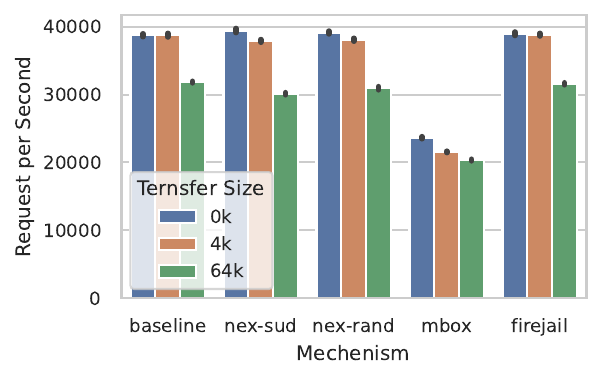}
    \caption{NGINX Benchmark [max. std. $<1.34\%$]}
    \label{fig:nginx}
\end{figure}
We evaluated a sandboxed NGINX server. The size of the request affects load characteristics, so we tested with file sizes of 0k, 4k, and 64k. 
All tests were run locally with NGINX workers set to 1.
HTTP requests are initiated using ab with a concurrency level of $64$. 
We use the request per second to calculate the overhead. 
The results of the evaluation are shown in Fig~\ref{fig:nginx}. 
In the case of using SUD, the maximum overhead caused by nexpoline is $5.3\%$. When using randomized seccomp, the maximum overhead is $2.8\%$. Firejail, implemented as a namespace sandbox, has an overhead of only $0.7\%$. In comparison, mbox, which uses seccomp and ptrace, has an average overhead of $40\%$. 
In summary, nexpoline offers a syscall-based sandbox with low overhead and opens up possibilities for other policies.

\begin{figure}[t]
    \centering
    
    \includegraphics[width=0.45\textwidth]{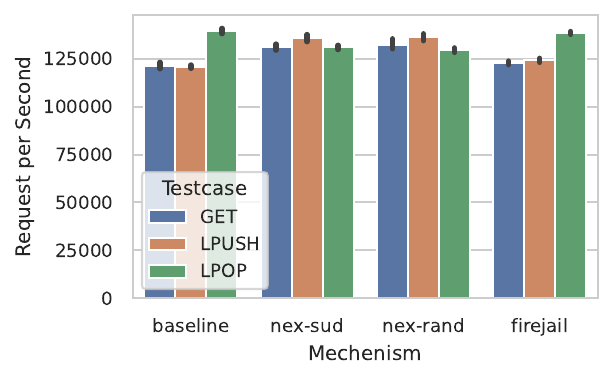}
    \caption{Redis Benchmark [max. std. $<3.50\%$]}
    \label{fig:redis}
\end{figure}

\textbf{Redis Database} -- We use the redis-benchmark for benchmark, and Fig~\ref{fig:redis} reports the results in requests per second. We used a concurrency level of 64. The test cases were GET, LPUSH, and LPOP, each with  different workload. 
The GET involves a simple key/value return, while LPUSH and LPOP involve operations on a list.
As mbox could not run Redis, only firejail is evaluated. 
The LPOP has the highest overhead, with \vsud, \vrnd having overheads of $6.1\%$, $7.1\%$, respectively. 
Firejail's maximum overhead is $0.8\%$, benefiting similarly from using only the sandboxing support provided by the kernel.

In conclusion, nexpoline provides nearly efficiency as firejail and enhances policy flexibility, enabling applications like gVisor and LibOSes, while significantly reducing overhead compared to seccomp based mbox.

\section{Known Limitation}
\textbf{io\_uring~\cite{Ringingi26:online} and vDSO} -- Not all system calls are based on \syscall. That is the case for io\_uring and vDSO.
They cannot be intercepted even for seccomp. Generally, they do not pose a security issue, but if necessary, disable vDSO and intercept the registration of io\_uring, especially considering that the performance gains from io\_uring are not clear~\cite{iouringi45:online}.
%
%
%

\textbf{rseq} -- Nexpoline needs to ensure that interrupts are controllable like signal, but rseq causes interrupts and its target address is not decide by the \syscall. 
We need to disable this system call. This is not a problem as rseq is not widely used. 

\textbf{ptrace and remote memory apis} -- ptrace can be used to debug child processes, thereby compromising the integrity of nexpoline in the child process, and therefore needs to be disabled. APIs related to remote memory access bypass the page protection, which circumvents MPK, thus need to be disabled.

\textbf{libc modification} -- Currently, nexpoline provides system call compatibility by modifying libc. For applications that do not use libc, it is necessary to modify the corresponding libraries to allow them to make system calls via \Code{syscall\_trap}, such as in the case of Golang. Alternatively, binary modification techniques, such as zpoline, can be used to improve compatibility in these situations.

\section{Related Work}
\textbf{System Call Interception} -- System Call Interception involves binary rewriting~\cite{zpoline, e9patch, pmemsysc0:online, xcontainer, Detours}, ptrace, seccomp, modifying libc~\cite{Graphenelibos, ERIM}, and etc~\cite{preload1}. Binary rewriting and modifying libc are inherently insecure and often require additional safeguards, such as seccomp, to mitigate risks like the subtle removal of risky syscall instructions embedded within other operations. Alternatively, ptrace is a usual method for system call interception but incurs significant performance overhead. In contrast, nexpoline uses binary rewriting or modifying libc purely for compatibility with existing applications, securing system calls through mechanisms like seccomp and SUD, which can be integrated with these modifications.

\textbf{Sandboxes} -- Sandboxes are used to prevent applications from performing illegal operations~\cite{mbox, firejail, li2014minibox, Watson:Capsicum:2010, Ford:Vx32:2008, seccomp-sandbox-chrome:online, bbh+2015, Kolosick22, nacl1, nacl2, nacl3}. 
It requires inspection and filtering of the application's system calls. 
LSM and namespaces are commonly used for sandboxing such as firejail~\cite{firejail} in which the kernel limits syscalls policies efficiently~\cite{firejail, seccomp-sandbox-chrome:online}, offers better performance~\cite{pmm+2012, LSMOverhead}. However, the limited and inflexible policies cannot be used for applications like gVisor~\cite{gVisor:online} or other LibOSes. For such techniques, ptrace or in-process isoaltion is the only alternative.

\textbf{In-process Isolation} --
MPK and other hardware extensions~\cite{intel_vol3, hfi, SecureCells},improve the performance of in-process isolation, it's use has become popular in research prototypes~\cite{Jenny, ShallNot, ERIM, Koning:No:2017, Litton:Lightweight:2016, Liu:Thwarting:2015, Park:libmpk:2019, Ghosn:Enclosure:2021, Brumley:Privtrans:2004a, Hodor, vdom, uswitch}. 
Due to MPK's limitations~\cite{ShallNot, Pitfalls}, they also need to restrict system calls. 
Previous work required modifications to the kernel and are unable to fully and securely support signals~\cite{Jenny, ShallNot}, thus limiting their usability and scope of application.
Nexpoline can be used as part of an in-process monitor to enhance their security and compatibility.

\textbf{Control-flow Integrity and Software Fault Isolation} -- Enforcing CFI~\cite{Intel:Control:2016, Wang:HyperSafe:2010, Ge:Fine:2016, Abadi:Control:2005, Budiu:Architectural:2006, 7054189, burow2017control, abadi2009control, mashtizadeh2015ccfi, christoulakis2016hcfi, criswell2014kcofi, Kuznetsov:Code:2014, Francillon:Defending:2009, Zeng:Combining:2011, SHARD, BASTION} or SFI~\cite{sfi1, Mao11SFI, Castro09SFI, Sehr10SFI, nacl1, nacl2, nacl3, NoNeedtoHide, Kolosick22} can also help intercept system calls, by restricting \syscall instructions to a specific range. And, they can thus ensures \syscall-s are not abused. But, both incur significant overhead compared to efficient hardware in-process isolation techniques due to tracking stack context and/or poor backward compatibility requiring recompilation.

\section{Conclusion}
This paper presents nexpoline, a mechanism that utilizes  MPK, Seccomp and SUD to make \syscall a privilege. It allows for exhaustive, safe, and low-overhead system call interception without requiring to modify the kernel. It can be used with other binary rewriting tools to enhance their security. Nexpoline provides secure and reliable support for system call policies in tools such as sandboxes, in-process monitors, and operating system emulation.
\newpage
\printbibliography

@inproceedings {Jenny,
    author = {David Schrammel and Samuel Weiser and Richard Sadek and Stefan Mangard},
    title = {Jenny: Securing Syscalls for {PKU-based} Memory Isolation Systems},
    booktitle = {31st USENIX Security Symposium (USENIX Security 22)},
    year = {2022},
    isbn = {978-1-939133-31-1},
    address = {Boston, MA},
    pages = {936--952},
    url = {https://www.usenix.org/conference/usenixsecurity22/presentation/schrammel},
    publisher = {USENIX Association},
    month = aug
}

@inproceedings{ShallNot,
    author = {Voulimeneas, Alexios and Vinck, Jonas and Mechelinck, Ruben and Volckaert, Stijn},
    title = {You Shall Not (by)Pass! Practical, Secure, and Fast PKU-Based Sandboxing},
    year = {2022},
    isbn = {9781450391627},
    publisher = {Association for Computing Machinery},
    address = {New York, NY, USA},
    url = {https://doi.org/10.1145/3492321.3519560},
    doi = {10.1145/3492321.3519560},
    booktitle = {Proceedings of the Seventeenth European Conference on Computer Systems},
    pages = {266–282},
    numpages = {17},
    location = {Rennes, France},
    series = {EuroSys '22}
}

@misc{PKUusage61Bug:online,
    author = {Kees Cook},
    title = {PKU usage improvements for threads - Kees Cook},
    howpublished = {\url{https://lore.kernel.org/lkml/202208221331.71C50A6F@keescook/}},
    month = {Aug},
    year = {2022},
    note = {(Accessed on 12/30/2023)}
}

@inproceedings{Ghosn:Enclosure:2021,
	title        = {Enclosure: Language-Based Restriction of Untrusted Libraries},
	author       = {Ghosn, Adrien and Kogias, Marios and Payer, Mathias and Larus, James R. and Bugnion, Edouard},
	year         = 2021,
	booktitle    = {Proceedings of the 26th ACM International Conference on Architectural Support for Programming Languages and Operating Systems},
	location     = {Virtual, USA},
	publisher    = {Association for Computing Machinery},
	address      = {New York, NY, USA},
	series       = {ASPLOS 2021},
	pages        = {255–267},
	doi          = {10.1145/3445814.3446728},
	isbn         = 9781450383172,
	url          = {https://doi.org/10.1145/3445814.3446728},
	numpages     = 13
}

@inproceedings{PKRU-safe,
author = {Kirth, Paul and Dickerson, Mitchel and Crane, Stephen and Larsen, Per and Dabrowski, Adrian and Gens, David and Na, Yeoul and Volckaert, Stijn and Franz, Michael},
title = {PKRU-Safe: Automatically Locking down the Heap between Safe and Unsafe Languages},
year = {2022},
isbn = {9781450391627},
publisher = {Association for Computing Machinery},
address = {New York, NY, USA},
url = {https://doi.org/10.1145/3492321.3519582},
doi = {10.1145/3492321.3519582},
booktitle = {Proceedings of the Seventeenth European Conference on Computer Systems},
pages = {132–148},
numpages = {17},
location = {Rennes, France},
series = {EuroSys '22}
}

@inproceedings {ERIM,
author = {Anjo Vahldiek-Oberwagner and Eslam Elnikety and Nuno O. Duarte and Michael Sammler and Peter Druschel and Deepak Garg},
title = {{ERIM}: Secure, Efficient In-process Isolation with Protection Keys ({{{{{MPK}}}}})},
booktitle = {28th USENIX Security Symposium (USENIX Security 19)},
year = {2019},
isbn = {978-1-939133-06-9},
address = {Santa Clara, CA},
pages = {1221--1238},
url = {https://www.usenix.org/conference/usenixsecurity19/presentation/vahldiek-oberwagner},
publisher = {USENIX Association},
month = aug
}

@inproceedings {Pitfalls,
author = {R. Joseph Connor and Tyler McDaniel and Jared M. Smith and Max Schuchard},
title = {{PKU} Pitfalls: Attacks on {PKU-based} Memory Isolation Systems},
booktitle = {29th USENIX Security Symposium (USENIX Security 20)},
year = {2020},
isbn = {978-1-939133-17-5},
pages = {1409--1426},
url = {https://www.usenix.org/conference/usenixsecurity20/presentation/connor},
publisher = {USENIX Association},
month = aug
}

@inproceedings{Koning:No:2017,
	title        = {No {{Need}} to {{Hide}}: {{Protecting Safe Regions}} on {{Commodity Hardware}}},
	shorttitle   = {No {{Need}} to {{Hide}}},
	author       = {Koning, Koen and Chen, Xi and Bos, Herbert and Giuffrida, Cristiano and Athanasopoulos, Elias},
	year         = 2017,
	booktitle    = {Proceedings of the {{Twelfth European Conference}} on {{Computer Systems}}},
	location     = {{New York, NY, USA}},
	publisher    = {{ACM}},
	address      = {New York, NY, USA},
	series       = {{{EuroSys}} '17},
	pages        = {437--452},
	doi          = {10.1145/3064176.3064217},
	isbn         = {978-1-4503-4938-3},
	url          = {http://doi.acm.org/10.1145/3064176.3064217},
	urldate      = {2017-04-25},
	date         = 2017
}

@inproceedings{Litton:Lightweight:2016,
	title        = {Light-Weight {{Contexts}}: {{An OS Abstraction}} for {{Safety}} and {{Performance}}},
	shorttitle   = {Light-Weight {{Contexts}}},
	author       = {Litton, James and Vahldiek-Oberwagner, Anjo and Elnikety, Eslam and Garg, Deepak and Bhattacharjee, Bobby and Druschel, Peter},
	year         = 2016,
	booktitle    = {Proceedings of the 12th {{USENIX Conference}} on {{Operating Systems Design}} and {{Implementation}}},
	location     = {{Berkeley, CA, USA}},
	publisher    = {{USENIX Association}},
	address      = {Berkeley, CA, USA},
	series       = {{{OSDI}}'16},
	pages        = {49--64},
	isbn         = {978-1-931971-33-1},
	url          = {http://dl.acm.org/citation.cfm?id=3026877.3026882},
	date         = 2016
}

@inproceedings{Liu:Thwarting:2015,
	title        = {Thwarting {{Memory Disclosure}} with {{Efficient Hypervisor}}-Enforced {{Intra}}-Domain {{Isolation}}},
	author       = {Liu, Yutao and Zhou, Tianyu and Chen, Kexin and Chen, Haibo and Xia, Yubin},
	year         = 2015,
	booktitle    = {Proceedings of the {{22Nd ACM SIGSAC Conference}} on {{Computer}} and {{Communications Security}}},
	location     = {{New York, NY, USA}},
	publisher    = {{ACM}},
	address      = {New York, NY, USA},
	series       = {{{CCS}} '15},
	pages        = {1607--1619},
	doi          = {10.1145/2810103.2813690},
	isbn         = {978-1-4503-3832-5},
	url          = {http://doi.acm.org/10.1145/2810103.2813690},
	urldate      = {2016-06-15},
	date         = 2015
}

@inproceedings{Park:libmpk:2019,
	title        = {Libmpk: {{Software Abstraction}} for {{Intel Memory Protection Keys}} ({{Intel}} {{MPK}})},
	shorttitle   = {Libmpk},
	author       = {Park, Soyeon and Lee, Sangho and Xu, Wen and Moon, Hyungon and Kim, Taesoo},
	pages        = {241--254},
	isbn         = {978-1-939133-03-8},
	url          = {https://www.usenix.org/conference/atc19/presentation/park-soyeon},
	urldate      = {2020-06-18},
	date         = 2019,
	eventtitle   = {2019 {{USENIX}} {{Annual Technical Conference}} ({{USENIX}} {{ATC}} 19)},
	langid       = {english}
}

@manual{intel_vol3,
  title        = "{Intel 64 and IA-32 Architectures Software Developer’s Manual, Volume 3}",
  author       = "{Intel Corporation}",
  organization = "{Intel Corporation}",
  address      = "{Santa Clara, California}",
  month        = "Sept",
  year         = "2023",
}

@INPROCEEDINGS{uswitch,
  author={Peng, Dinglan and Liu, Congyu and Palit, Tapti and Fonseca, Pedro and Vahldiek-Oberwagner, Anjo and Vij, Mona},
  booktitle={2023 IEEE Symposium on Security and Privacy (SP)}, 
  title={$\mu$Switch: Fast Kernel Context Isolation with Implicit Context Switches}, 
  year={2023},
  volume={},
  number={},
  pages={2956-2973},
  doi={10.1109/SP46215.2023.10179284}
}

@inproceedings{vdom,
    author = {Yuan, Ziqi and Hong, Siyu and Chang, Rui and Zhou, Yajin and Shen, Wenbo and Ren, Kui},
    title = {VDom: Fast and Unlimited Virtual Domains on Multiple Architectures},
    year = {2023},
    isbn = {9781450399166},
    publisher = {Association for Computing Machinery},
    address = {New York, NY, USA},
    url = {https://doi.org/10.1145/3575693.3575735},
    doi = {10.1145/3575693.3575735},
    booktitle = {Proceedings of the 28th ACM International Conference on Architectural Support for Programming Languages and Operating Systems, Volume 2},
    pages = {905–919},
    numpages = {15},
    location = {Vancouver, BC, Canada},
    series = {ASPLOS 2023}
}

@inproceedings {Hodor,
    author = {Mohammad Hedayati and Spyridoula Gravani and Ethan Johnson and John Criswell and Michael L. Scott and Kai Shen and Mike Marty},
    title = {Hodor: {Intra-Process} Isolation for {High-Throughput} Data Plane Libraries},
    booktitle = {2019 USENIX Annual Technical Conference (USENIX ATC 19)},
    year = {2019},
    isbn = {978-1-939133-03-8},
    address = {Renton, WA},
    pages = {489--504},
    url = {http://www.usenix.org/conference/atc19/presentation/hedayati-hodor},
    publisher = {USENIX Association},
    month = jul
}

@inproceedings{hfi,
author = {Narayan, Shravan and Garfinkel, Tal and Taram, Mohammadkazem and Rudek, Joey and Moghimi, Daniel and Johnson, Evan and Fallin, Chris and Vahldiek-Oberwagner, Anjo and LeMay, Michael and Sahita, Ravi and Tullsen, Dean and Stefan, Deian},
title = {Going beyond the Limits of SFI: Flexible and Secure Hardware-Assisted In-Process Isolation with HFI},
year = {2023},
isbn = {9781450399180},
publisher = {Association for Computing Machinery},
address = {New York, NY, USA},
url = {https://doi.org/10.1145/3582016.3582023},
doi = {10.1145/3582016.3582023},
booktitle = {Proceedings of the 28th ACM International Conference on Architectural Support for Programming Languages and Operating Systems, Volume 3},
pages = {266–281},
numpages = {16},
location = {Vancouver, BC, Canada},
series = {ASPLOS 2023}
}

@INPROCEEDINGS{SecureCells,
  author={Bhattacharyya, Atri and Hofhammer, Florian and Li, Yuanlong and Gupta, Siddharth and Sanchez, Andres and Falsafi, Babak and Payer, Mathias},
  booktitle={2023 IEEE Symposium on Security and Privacy (SP)}, 
  title={SecureCells: A Secure Compartmentalized Architecture}, 
  year={2023},
  volume={},
  number={},
  pages={2921-2939},
  doi={10.1109/SP46215.2023.10179472}}

@inproceedings{e9patch,
    author = {Duck, Gregory J. and Gao, Xiang and Roychoudhury, Abhik},
    title = {Binary Rewriting without Control Flow Recovery},
    year = {2020},
    isbn = {9781450376136},
    publisher = {Association for Computing Machinery},
    address = {New York, NY, USA},
    url = {https://doi.org/10.1145/3385412.3385972},
    doi = {10.1145/3385412.3385972},
    booktitle = {Proceedings of the 41st ACM SIGPLAN Conference on Programming Language Design and Implementation},
    pages = {151–163},
    numpages = {13},
    location = {London, UK},
    series = {PLDI 2020}
}

@misc{pmemsysc0:online,
    author = {Persistent Memory Programming},
    title = {pmem/syscall\_intercept: The system call intercepting library},
    howpublished = {\url{https://github.com/pmem/syscall_intercept}},
    month = {},
    year = {2022},
    note = {(Accessed on 12/21/2023)}
}

@inproceedings {zpoline,
    author = {Kenichi Yasukata and Hajime Tazaki and Pierre-Louis Aublin and Kenta Ishiguro},
    title = {zpoline: a system call hook mechanism based on binary rewriting},
    booktitle = {2023 USENIX Annual Technical Conference (USENIX ATC 23)},
    year = {2023},
    isbn = {978-1-939133-35-9},
    address = {Boston, MA},
    pages = {293--300},
    url = {https://www.usenix.org/conference/atc23/presentation/yasukata},
    publisher = {USENIX Association},
    month = jul
}

@inproceedings{xcontainer,
author = {Shen, Zhiming and Sun, Zhen and Sela, Gur-Eyal and Bagdasaryan, Eugene and Delimitrou, Christina and Van Renesse, Robbert and Weatherspoon, Hakim},
title = {X-Containers: Breaking Down Barriers to Improve Performance and Isolation of Cloud-Native Containers},
year = {2019},
isbn = {9781450362405},
publisher = {Association for Computing Machinery},
address = {New York, NY, USA},
url = {https://doi.org/10.1145/3297858.3304016},
doi = {10.1145/3297858.3304016},
booktitle = {Proceedings of the Twenty-Fourth International Conference on Architectural Support for Programming Languages and Operating Systems},
pages = {121–135},
numpages = {15},
location = {Providence, RI, USA},
series = {ASPLOS '19}
}

@article{preload1,
author = {Chamith, Buddhika and Svensson, Bo Joel and Dalessandro, Luke and Newton, Ryan R.},
title = {Instruction Punning: Lightweight Instrumentation for X86-64},
year = {2017},
issue_date = {June 2017},
publisher = {Association for Computing Machinery},
address = {New York, NY, USA},
volume = {52},
number = {6},
issn = {0362-1340},
url = {https://doi.org/10.1145/3140587.3062344},
doi = {10.1145/3140587.3062344},
journal = {SIGPLAN Not.},
month = {jun},
pages = {320–332},
numpages = {13}
}

@inproceedings{Detours,
author = {Hunt, Galen and Brubacher, Doug},
title = {Detours: Binary Interception of Win32 Functions},
year = {1999},
publisher = {USENIX Association},
address = {USA},
booktitle = {Proceedings of the 3rd Conference on USENIX Windows NT Symposium - Volume 3},
pages = {14},
numpages = {1},
location = {Seattle, Washington},
series = {WINSYM'99}
}

@misc{Improvin37:online,
    author = {byValentin Rothberg},
    title = {Improving Linux container security with seccomp | Enable Sysadmin},
    howpublished = {\url{https://www.redhat.com/sysadmin/container-security-seccomp}},
    month = {},
    year = {2020},
    note = {(Accessed on 12/21/2023)}
}

@misc{Seccomps33:online,
    author = {Docker},
    title = {Seccomp security profiles for Docker | Docker Docs},
    howpublished = {\url{https://docs.docker.com/engine/security/seccomp/}},
    month = {},
    year = {2016},
    note = {(Accessed on 12/21/2023)}
}

@misc{seccomp-sandbox-chrome:online,
    author = {Julien Tinnes},
    title = {cr0 blog: Introducing Chrome's next-generation Linux sandbox},
    howpublished = {\url{https://blog.cr0.org/2012/09/introducing-chromes-next-generation.html}},
    month = {},
    year = {2012},
    note = {(Accessed on 12/21/2023)}
}

@inproceedings {Graphenelibos,
    author = {Chia-che Tsai and Donald E. Porter and Mona Vij},
    title = {{Graphene-SGX}: A Practical Library {OS} for Unmodified Applications on {SGX}},
    booktitle = {2017 USENIX Annual Technical Conference (USENIX ATC 17)},
    year = {2017},
    isbn = {978-1-931971-38-6},
    address = {Santa Clara, CA},
    pages = {645--658},
    url = {https://www.usenix.org/conference/atc17/technical-sessions/presentation/tsai},
    publisher = {USENIX Association},
    month = jul
}

@misc{WineHQRu68:online,
    author = {Wine},
    title = {WineHQ - Run Windows applications on Linux, BSD, Solaris and macOS},
    howpublished = {\url{https://www.winehq.org/}},
    month = {},
    year = {2023},
    note = {(Accessed on 12/21/2023)}
}

@inproceedings{Unikernels,
    author = {Madhavapeddy, Anil and Mortier, Richard and Rotsos, Charalampos and Scott, David and Singh, Balraj and Gazagnaire, Thomas and Smith, Steven and Hand, Steven and Crowcroft, Jon},
    title = {Unikernels: Library Operating Systems for the Cloud},
    year = {2013},
    isbn = {9781450318709},
    publisher = {Association for Computing Machinery},
    address = {New York, NY, USA},
    url = {https://doi.org/10.1145/2451116.2451167},
    doi = {10.1145/2451116.2451167},
    booktitle = {Proceedings of the Eighteenth International Conference on Architectural Support for Programming Languages and Operating Systems},
    pages = {461–472},
    numpages = {12},
    location = {Houston, Texas, USA},
    series = {ASPLOS '13}
}

@inproceedings {Dune,
    author = {Adam Belay and Andrea Bittau and Ali Mashtizadeh and David Terei and David Mazi{\`e}res and Christos Kozyrakis},
    title = {Dune: Safe User-level Access to Privileged {CPU} Features},
    booktitle = {10th USENIX Symposium on Operating Systems Design and Implementation (OSDI 12)},
    year = {2012},
    isbn = {978-1-931971-96-6},
    address = {Hollywood, CA},
    pages = {335--348},
    url = {https://www.usenix.org/conference/osdi12/technical-sessions/presentation/belay},
    publisher = {USENIX Association},
    month = oct
}

@inproceedings{Unikraft,
    author = {Kuenzer, Simon and B\u{a}doiu, Vlad-Andrei and Lefeuvre, Hugo and Santhanam, Sharan and Jung, Alexander and Gain, Gaulthier and Soldani, Cyril and Lupu, Costin and Teodorescu, \c{S}tefan and R\u{a}ducanu, Costi and Banu, Cristian and Mathy, Laurent and Deaconescu, R\u{a}zvan and Raiciu, Costin and Huici, Felipe},
    title = {Unikraft: Fast, Specialized Unikernels the Easy Way},
    year = {2021},
    isbn = {9781450383349},
    publisher = {Association for Computing Machinery},
    address = {New York, NY, USA},
    url = {https://doi.org/10.1145/3447786.3456248},
    doi = {10.1145/3447786.3456248},
    booktitle = {Proceedings of the Sixteenth European Conference on Computer Systems},
    pages = {376–394},
    numpages = {19},
    location = {Online Event, United Kingdom},
    series = {EuroSys '21}
}

@inproceedings{gVisorCost,
    author = {Young, Ethan G. and Zhu, Pengfei and Caraza-Harter, Tyler and Arpaci-Dusseau, Andrea C. and Arpaci-Dusseau, Remzi H.},
    title = {The True Cost of Containing: A GVisor Case Study},
    year = {2019},
    publisher = {USENIX Association},
    address = {USA},
    booktitle = {Proceedings of the 11th USENIX Conference on Hot Topics in Cloud Computing},
    pages = {16},
    numpages = {1},
    location = {Renton, WA, USA},
    series = {HotCloud'19}
}

@misc{gVisor:online,
    author = {Google Inc},
    title = {The Container Security Platform | gVisor},
    howpublished = {\url{https://gvisor.dev/}},
    month = {},
    year = {2023},
    note = {(Accessed on 12/21/2023)}
}

@misc{chromebug:online,
    author = {Igor Sak-Sakovskii},
    title = {1458911 - Security: Libxslt arbitrary file reading using document() method and external entities. - chromium},
    howpublished = {\url{https://bugs.chromium.org/p/chromium/issues/detail?id=1458911}},
    month = {Jun},
    year = {2023},
    note = {(Accessed on 12/21/2023)}
}

@misc{k8sseccomp:online,
    author = {Kubernetes},
    title = {Restrict a Container's Syscalls with seccomp | Kubernetes},
    howpublished = {\url{https://kubernetes.io/docs/tutorials/security/seccomp/}},
    month = {},
    year = {2023},
    note = {(Accessed on 12/21/2023)}
}

@inproceedings{qemu,
    author = {Bellard, Fabrice},
    title = {QEMU, a Fast and Portable Dynamic Translator},
    year = {2005},
    publisher = {USENIX Association},
    address = {USA},
    booktitle = {Proceedings of the Annual Conference on USENIX Annual Technical Conference},
    pages = {41},
    numpages = {1},
    location = {Anaheim, CA},
    series = {ATEC '05}
}

@inproceedings {mbox,
    author = {Taesoo Kim and Nickolai Zeldovich},
    title = {Practical and Effective Sandboxing for Non-root Users},
    booktitle = {2013 USENIX Annual Technical Conference (USENIX ATC 13)},
    year = {2013},
    isbn = {978-1-931971-01-0},
    address = {San Jose, CA},
    pages = {139--144},
    url = {https://www.usenix.org/conference/atc13/technical-sessions/presentation/kim},
    publisher = {USENIX Association},
    month = jun
}

@misc{Seccomp,
    author = {Will Drewry},
    title = {dynamic seccomp policies (using BPF filters) [LWN.net]},
    howpublished = {\url{https://lwn.net/Articles/475019/}},
    month = {},
    year = {2012},
    note = {(Accessed on 12/21/2023)}
}

@misc{SUD:online,
    author = {Linux},
    title = {Syscall User Dispatch — The Linux Kernel documentation},
    howpublished = {\url{https://docs.kernel.org/admin-guide/syscall-user-dispatch.html}},
    month = {},
    year = {2023},
    note = {(Accessed on 12/21/2023)}
}

@misc{iouringi45:online,
    author = {},
    title = {io\_uring is slower than epoll · Issue \#189 · axboe/liburing},
    howpublished = {\url{https://github.com/axboe/liburing/issues/189}},
    month = {},
    year = {2020},
    note = {(Accessed on 12/21/2023)}
}

@misc{Ringingi26:online,
    author = {Jonathan Corbet},
    title = {Ringing in a new asynchronous I/O API [LWN.net]},
    howpublished = {\url{https://lwn.net/Articles/776703/}},
    month = {Jan},
    year = {2019},
    note = {(Accessed on 12/23/2023)}
}

@misc{darlingh95:online,
    author = {darlinghq},
    title = {darlinghq/darling: Darwin/macOS emulation layer for Linux},
    howpublished = {\url{https://github.com/darlinghq/darling}},
    month = {},
    year = {2023},
    note = {(Accessed on 12/22/2023)}
}

@misc{RFCPATCH40:online,
author = {Michael Sammler},
title = {[RFC PATCH] seccomp: Add protection keys into seccomp\_data - Michael Sammler},
howpublished = {\url{https://lore.kernel.org/linux-api/20181029112343.27454-1-msammler@mpi-sws.org/}},
month = {Oct},
year = {2018},
note = {(Accessed on 01/09/2024)}
}

@misc{firejail,
author = {Firejail},
title = {Firejail | security sandbox},
howpublished = {\url{https://firejail.wordpress.com/}},
month = {Oct},
year = {2023},
note = {(Accessed on 01/10/2024)}
}

@book{selinux,
author = {McCarty, Bill},
title = {SELinux: NSA's Open Source Security Enhanced Linux},
year = {2004},
isbn = {0596007167},
publisher = {O'Reilly Media, Inc.}
}

@inproceedings{LSMOverhead,
author = {Zhang, Wenhui and Liu, Peng and Jaeger, Trent},
title = {Analyzing the Overhead of File Protection by Linux Security Modules},
year = {2021},
isbn = {9781450382878},
publisher = {Association for Computing Machinery},
address = {New York, NY, USA},
url = {https://doi.org/10.1145/3433210.3453078},
doi = {10.1145/3433210.3453078},
booktitle = {Proceedings of the 2021 ACM Asia Conference on Computer and Communications Security},
pages = {393–406},
numpages = {14},
location = {Virtual Event, Hong Kong},
series = {ASIA CCS '21}
}

@misc{landlock,
author = {Mickaël Salaün},
title = {Landlock: unprivileged access control — The Linux Kernel documentation},
howpublished = {\url{https://docs.kernel.org/userspace-api/landlock.html}},
month = {Oct},
year = {2023},
note = {(Accessed on 01/15/2024)}
}

@inproceedings{LinuxMAC,
author = {Brimhall, Brennon and Garrard, Justin and De La Garza, Christopher and Coffman, Joel},
title = {A Comparative Analysis of Linux Mandatory Access Control Policy Enforcement Mechanisms},
year = {2023},
isbn = {9798400700859},
publisher = {Association for Computing Machinery},
address = {New York, NY, USA},
url = {https://doi.org/10.1145/3578357.3589454},
doi = {10.1145/3578357.3589454},
booktitle = {Proceedings of the 16th European Workshop on System Security},
pages = {1–7},
numpages = {7},
location = {Rome, Italy},
series = {EUROSEC '23}
}

@INPROCEEDINGS{AppArmorP,
  author={Ecarot, Thibaud and Dussault, Samuel and Souid, Ameni and Lavoie, Luc and Ethier, Jean-François},
  booktitle={2020 7th International Conference on Internet of Things: Systems, Management and Security (IOTSMS)}, 
  title={AppArmor For Health Data Access Control: Assessing Risks and Benefits}, 
  year={2020},
  volume={},
  number={},
  pages={1-7},
  doi={10.1109/IOTSMS52051.2020.9340206}}

@misc{AppArmor,
author = {},
title = {AppArmor},
howpublished = {\url{https://apparmor.net/}},
month = {July},
year = {2023},
note = {(Accessed on 01/15/2024)}
}

@misc{LSM,
author = {},
title = {Linux Security Module Usage — The Linux Kernel documentation},
howpublished = {\url{https://docs.kernel.org/admin-guide/LSM/index.html}},
month = {},
year = {},
note = {(Accessed on 01/15/2024)}
}

@misc{namespace,
author = {},
title = {namespaces(7) - Linux manual page},
howpublished = {\url{https://man7.org/linux/man-pages/man7/namespaces.7.html}},
month = {},
year = {},
note = {(Accessed on 01/15/2024)}
}

@misc{cgroups,
author = {Tejun Heo},
title = {Control Group v2 — The Linux Kernel documentation},
howpublished = {\url{https://www.kernel.org/doc/html/latest/admin-guide/cgroup-v2.html}},
month = {Oct},
year = {2015},
note = {(Accessed on 01/15/2024)}
}

@misc{patchSUD,
author = {Gregory Price},
title = {[PATCH v15 0/4] Checkpoint Support for Syscall User Dispatch},
howpublished = {\url{https://lore.kernel.org/lkml/ZCYP+4gRZDqC0lRo@arm.com/T/}},
month = {March},
year = {2023},
note = {(Accessed on 01/15/2024)}
}

@inproceedings{pmm+2012,
	title        = {{Hi-Fi: Collecting High-Fidelity Whole-System Provenance}},
	author       = {Pohly, D.J. and McLaughlin, S. and McDaniel, P. and Butler, K.},
	year         = 2012,
	booktitle    = {Proceedings of the 2012 Annual Computer Security Applications Conference},
	address      = {Orlando, FL, USA},
	series       = {ACSAC '12}
}

@inproceedings{li2014minibox,
  title={MiniBox: A Two-Way Sandbox for x86 Native Code},
  author={Li, Yanlin and McCune, Jonathan and Newsome, James and Perrig, Adrian and Baker, Brandon and Drewry, Will},
  booktitle={2014 USENIX annual technical conference (USENIX ATC 14)},
  pages={409--420},
  year={2014}
}

@inproceedings{Watson:Capsicum:2010,
	title        = {Capsicum: {{Practical Capabilities}} for {{UNIX}}},
	shorttitle   = {Capsicum},
	author       = {Watson, Robert N. M. and Anderson, Jonathan and Laurie, Ben and Kennaway, Kris},
	year         = 2010,
	booktitle    = {Proceedings of the 19th {{USENIX Conference}} on {{Security}}},
	location     = {{Berkeley, CA, USA}},
	publisher    = {{USENIX Association}},
	address      = {Berkeley, CA, USA},
	series       = {USENIX Security'10},
	volume       = 46,
	pages        = {3--3},
	isbn         = {888-7-6666-5555-4},
	url          = {http://dl.acm.org/citation.cfm?id=1929820.1929824},
	urldate      = {2015-01-14},
	date         = 2010
}

@inproceedings{Brumley:Privtrans:2004a,
	title        = {Privtrans: {{Automatically Partitioning Programs}} for {{Privilege Separation}}},
	shorttitle   = {Privtrans},
	author       = {Brumley, David and Song, Dawn},
	booktitle    = {Proceedings of the 13th {{Conference}} on {{USENIX Security Symposium}} - {{Volume}} 13},
	location     = {{Berkeley, CA, USA}},
	publisher    = {{USENIX Association}},
	series       = {{{SSYM}}'04},
	pages        = {5--5},
	url          = {http://dl.acm.org/citation.cfm?id=1251375.1251380},
	urldate      = {2015-05-15},
	date         = 2004
}

@inproceedings{Ford:Vx32:2008,
	title        = {Vx32: {{Lightweight User}}-Level {{Sandboxing}} on the X86},
	shorttitle   = {Vx32},
	author       = {Ford, Bryan and Cox, Russ},
	booktitle    = {{{USENIX}} 2008 {{Annual Technical Conference}}},
	location     = {{Berkeley, CA, USA}},
	publisher    = {{USENIX Association}},
	series       = {{{ATC}}'08},
	pages        = {293--306},
	url          = {http://dl.acm.org/citation.cfm?id=1404014.1404039},
	urldate      = {2015-05-12},
	date         = 2008
}

@inproceedings{bbh+2015,
	title        = {Boxify: Full-fledged App Sandboxing for Stock Android},
	author       = {Backes, Michael and Bugiel, Sven and Hammer, Christian and Schranz, Oliver and Von Styp-Rekowsky, Philipp},
	year         = 2015,
	booktitle    = {Proceedings of the 24th USENIX Conference on Security Symposium},
	location     = {Washington, D.C.},
	publisher    = {USENIX Association},
	address      = {Berkeley, CA, USA},
	series       = {SEC'15},
	pages        = {691--706},
	isbn         = {978-1-931971-232},
	url          = {http://dl.acm.org/citation.cfm?id=2831143.2831187},
	acmid        = 2831187,
	numpages     = 16
}

@inproceedings{Lei2017SPEAKERSE,
  title={SPEAKER: Split-Phase Execution of Application Containers},
  author={Lingguang Lei and Jianhua Sun and Kun Sun and Chris Shenefiel and Rui Ma and Yuewu Wang and Qi Li},
  booktitle={International Conference on Detection of intrusions and malware, and vulnerability assessment},
  year={2017},
  url={https://api.semanticscholar.org/CorpusID:20448694}
}

@INPROCEEDINGS{mine,
  author={Wan, Zhiyuan and Lo, David and Xia, Xin and Cai, Liang and Li, Shanping},
  booktitle={2017 IEEE International Conference on Software Testing, Verification and Validation (ICST)}, 
  title={Mining Sandboxes for Linux Containers}, 
  year={2017},
  volume={},
  number={},
  pages={92-102},
  doi={10.1109/ICST.2017.16}}

@inproceedings {sysfilter,
author = {Nicholas DeMarinis and Kent Williams-King and Di Jin and Rodrigo Fonseca and Vasileios P. Kemerlis},
title = {sysfilter: Automated System Call Filtering for Commodity Software},
booktitle = {23rd International Symposium on Research in Attacks, Intrusions and Defenses (RAID 2020)},
year = {2020},
isbn = {978-1-939133-18-2},
address = {San Sebastian},
pages = {459--474},
url = {https://www.usenix.org/conference/raid2020/presentation/demarinis},
publisher = {USENIX Association},
month = oct
}

@article{automodel,
author = {Pailoor, Shankara and Wang, Xinyu and Shacham, Hovav and Dillig, Isil},
title = {Automated Policy Synthesis for System Call Sandboxing},
year = {2020},
issue_date = {November 2020},
publisher = {Association for Computing Machinery},
address = {New York, NY, USA},
volume = {4},
number = {OOPSLA},
url = {https://doi.org/10.1145/3428203},
doi = {10.1145/3428203},
journal = {Proc. ACM Program. Lang.},
month = {nov},
articleno = {135},
numpages = {26}
}

@inproceedings {confine,
author = {Seyedhamed Ghavamnia and Tapti Palit and Azzedine Benameur and Michalis Polychronakis},
title = {Confine: Automated System Call Policy Generation for Container Attack Surface Reduction},
booktitle = {23rd International Symposium on Research in Attacks, Intrusions and Defenses (RAID 2020)},
year = {2020},
isbn = {978-1-939133-18-2},
address = {San Sebastian},
pages = {443--458},
url = {https://www.usenix.org/conference/raid2020/presentation/ghavanmnia},
publisher = {USENIX Association},
month = oct
}

@inproceedings {temporal,
author = {Seyedhamed Ghavamnia and Tapti Palit and Shachee Mishra and Michalis Polychronakis},
title = {Temporal System Call Specialization for Attack Surface Reduction},
booktitle = {29th USENIX Security Symposium (USENIX Security 20)},
year = {2020},
isbn = {978-1-939133-17-5},
pages = {1749--1766},
url = {https://www.usenix.org/conference/usenixsecurity20/presentation/ghavamnia},
publisher = {USENIX Association},
month = aug
}

@article{sfi1,
author = {Wahbe, Robert and Lucco, Steven and Anderson, Thomas E. and Graham, Susan L.},
title = {Efficient Software-Based Fault Isolation},
year = {1993},
issue_date = {Dec. 1993},
publisher = {Association for Computing Machinery},
address = {New York, NY, USA},
volume = {27},
number = {5},
issn = {0163-5980},
url = {https://doi.org/10.1145/173668.168635},
doi = {10.1145/173668.168635},
journal = {SIGOPS Oper. Syst. Rev.},
month = {dec},
pages = {203–216},
numpages = {14}
}

@inproceedings{Mao11SFI,
	title        = {Software Fault Isolation with API Integrity and Multi-Principal Modules},
	author       = {Mao, Yandong and Chen, Haogang and Zhou, Dong and Wang, Xi and Zeldovich, Nickolai and Kaashoek, M. Frans},
	year         = 2011,
	booktitle    = {Proceedings of the Twenty-Third ACM Symposium on Operating Systems Principles},
	location     = {Cascais, Portugal},
	publisher    = {Association for Computing Machinery},
	address      = {New York, NY, USA},
	series       = {SOSP '11},
	pages        = {115–128},
	doi          = {10.1145/2043556.2043568},
	isbn         = 9781450309776,
	url          = {https://doi.org/10.1145/2043556.2043568},
	numpages     = 14
}

@inproceedings{Castro09SFI,
	title        = {Fast Byte-Granularity Software Fault Isolation},
	author       = {Castro, Miguel and Costa, Manuel and Martin, Jean-Philippe and Peinado, Marcus and Akritidis, Periklis and Donnelly, Austin and Barham, Paul and Black, Richard},
	year         = 2009,
	booktitle    = {Proceedings of the ACM SIGOPS 22nd Symposium on Operating Systems Principles},
	location     = {Big Sky, Montana, USA},
	publisher    = {Association for Computing Machinery},
	address      = {New York, NY, USA},
	series       = {SOSP '09},
	pages        = {45–58},
	doi          = {10.1145/1629575.1629581},
	isbn         = 9781605587523,
	url          = {https://doi.org/10.1145/1629575.1629581},
	numpages     = 14
}

@inproceedings{Sehr10SFI,
	title        = {Adapting Software Fault Isolation to Contemporary CPU Architectures},
	author       = {Sehr, David and Muth, Robert and Biffle, Cliff and Khimenko, Victor and Pasko, Egor and Schimpf, Karl and Yee, Bennet and Chen, Brad},
	year         = 2010,
	booktitle    = {Proceedings of the 19th USENIX Conference on Security},
	location     = {Washington, DC},
	publisher    = {USENIX Association},
	address      = {USA},
	series       = {USENIX Security'10},
	pages        = 1,
	isbn         = 8887666655554,
	numpages     = 1
}

@inbook{Intel:Control:2016,
	title        = {{Chapter 18 Control-flow enforcement technology (CET)}},
	author       = {{Intel}},
	year         = 2021,
	month        = jun,
	booktitle    = {Intel® 64 and IA-32 Architectures Software Developer’s Manual},
	publisher    = {Intel},
	volume       = 1,
	number       = {334525-001},
	pages        = {409--422},
	url          = {software.intel.com/content/www/cn/zh/develop/articles/intel-sdm.html},
	type         = {White {{Paper}}},
	date         = {2016-06}
}

@inproceedings{Wang:HyperSafe:2010,
	title        = {{{HyperSafe}}: {{A Lightweight Approach}} to {{Provide Lifetime Hypervisor Control}}-{{Flow Integrity}}},
	shorttitle   = {{{HyperSafe}}},
	author       = {Wang, Zhi and Jiang, Xuxian},
	year         = 2010,
	booktitle    = {Proceedings of the 2010 {{IEEE Symposium}} on {{Security}} and {{Privacy}}},
	location     = {{Washington, DC, USA}},
	publisher    = {{IEEE Computer Society}},
	address      = {Washington, DC, USA},
	series       = {SP '10},
	pages        = {380--395},
	doi          = {10.1109/SP.2010.30},
	isbn         = {978-0-7695-4035-1},
	url          = {http://dx.doi.org/10.1109/SP.2010.30},
	urldate      = {2015-01-14},
	date         = 2010
}

@inproceedings{Ge:Fine:2016,
	title        = {Fine-{{Grained Control}}-{{Flow Integrity}} for {{Kernel Software}}},
	author       = {Ge, X. and Talele, N. and Payer, M. and Jaeger, T.},
	year         = 2016,
	month        = mar,
	booktitle    = {2016 {{IEEE European Symposium}} on {{Security}} and {{Privacy}} ({{EuroS P}})},
	pages        = {179--194},
	doi          = {10.1109/EuroSP.2016.24},
	date         = {2016-03},
	eventtitle   = {2016 {{IEEE European Symposium}} on {{Security}} and {{Privacy}} ({{EuroS P}})}
}

@inproceedings{Abadi:Control:2005,
	title        = {Control-Flow Integrity},
	author       = {Abadi, M. and Budiu, M. and Erlingsson, Ú. and Ligatti, J.},
	year         = 2005,
	booktitle    = {Proceedings of the 12th {{ACM}} Conference on {{Computer}} and Communications Security},
	pages        = {340--353},
	date         = 2005,
	organization = {ACM}
}

@inproceedings{Budiu:Architectural:2006,
	title        = {Architectural {{Support}} for {{Software}}-Based {{Protection}}},
	author       = {Budiu, Mihai and Erlingsson, Úlfar and Abadi, Martín},
	booktitle    = {Proceedings of the 1st {{Workshop}} on {{Architectural}} and {{System Support}} for {{Improving Software Dependability}}},
	location     = {{New York, NY, USA}},
	publisher    = {{ACM}},
	series       = {{{ASID}} '06},
	pages        = {42--51},
	doi          = {10.1145/1181309.1181316},
	isbn         = {1-59593-576-2},
	url          = {http://doi.acm.org/10.1145/1181309.1181316},
	urldate      = {2014-09-23},
	date         = 2006
}

@INPROCEEDINGS{7054189,
  author={Arthur, William and Mehne, Ben and Das, Reetuparna and Austin, Todd},
  booktitle={2015 IEEE/ACM International Symposium on Code Generation and Optimization (CGO)}, 
  title={Getting in control of your control flow with control-data isolation}, 
  year={2015},
  volume={},
  number={},
  pages={79-90},
  doi={10.1109/CGO.2015.7054189}}

@article{burow2017control,
  title={Control-flow integrity: Precision, security, and performance},
  author={Burow, Nathan and Carr, Scott A and Nash, Joseph and Larsen, Per and Franz, Michael and Brunthaler, Stefan and Payer, Mathias},
  journal={ACM Computing Surveys (CSUR)},
  volume={50},
  number={1},
  pages={1--33},
  year={2017},
  publisher={ACM New York, NY, USA}
}

@article{abadi2009control,
  title={Control-flow integrity principles, implementations, and applications},
  author={Abadi, Mart{\'\i}n and Budiu, Mihai and Erlingsson, Ulfar and Ligatti, Jay},
  journal={ACM Transactions on Information and System Security (TISSEC)},
  volume={13},
  number={1},
  pages={1--40},
  year={2009},
  publisher={ACM New York, NY, USA}
}

@inproceedings{mashtizadeh2015ccfi,
  title={CCFI: Cryptographically enforced control flow integrity},
  author={Mashtizadeh, Ali Jose and Bittau, Andrea and Boneh, Dan and Mazi{\`e}res, David},
  booktitle={Proceedings of the 22nd ACM SIGSAC Conference on Computer and Communications Security},
  pages={941--951},
  year={2015}
}

@inproceedings{christoulakis2016hcfi,
  title={HCFI: Hardware-enforced control-flow integrity},
  author={Christoulakis, Nick and Christou, George and Athanasopoulos, Elias and Ioannidis, Sotiris},
  booktitle={Proceedings of the Sixth ACM Conference on Data and Application Security and Privacy},
  pages={38--49},
  year={2016}
}

@inproceedings{criswell2014kcofi,
  title={KCoFI: Complete control-flow integrity for commodity operating system kernels},
  author={Criswell, John and Dautenhahn, Nathan and Adve, Vikram},
  booktitle={2014 IEEE symposium on security and privacy},
  pages={292--307},
  year={2014},
  organization={IEEE}
}

@inproceedings{Kuznetsov:Code:2014,
	title        = {Code-{{Pointer Integrity}}},
	author       = {Kuznetsov, Volodymyr and Szekeres, Laszlo and Payer, Mathias and Candea, George and Sekar, R. and Song, Dawn},
	booktitle    = {11th {{USENIX Symposium}} on {{Operating Systems Design}} and {{Implementation}} ({{OSDI}} 14)},
	location     = {{Broomfield, CO}},
	publisher    = {{USENIX Association}},
	pages        = {147--163},
	isbn         = {978-1-931971-16-4},
	url          = {https://www.usenix.org/conference/osdi14/technical-sessions/presentation/kuznetsov},
	date         = {2014-10}
}

@inproceedings{Francillon:Defending:2009,
	title        = {Defending {{Embedded Systems Against Control Flow Attacks}}},
	author       = {Francillon, Aurélien and Perito, Daniele and Castelluccia, Claude},
	booktitle    = {Proceedings of the {{First ACM Workshop}} on {{Secure Execution}} of {{Untrusted Code}}},
	location     = {{New York, NY, USA}},
	publisher    = {{ACM}},
	series       = {{{SecuCode}} '09},
	pages        = {19--26},
	doi          = {10.1145/1655077.1655083},
	isbn         = {978-1-60558-782-0},
	url          = {http://doi.acm.org/10.1145/1655077.1655083},
	urldate      = {2014-09-04},
	date         = 2009
}

@inproceedings{Zeng:Combining:2011,
	title        = {Combining Control-Flow Integrity and Static Analysis for Efficient and Validated Data Sandboxing},
	author       = {Zeng, Bin and Tan, Gang and Morrisett, Greg},
	booktitle    = {Proceedings of the 18th {{ACM}} Conference on {{Computer}} and Communications Security},
	location     = {{New York, NY, USA}},
	publisher    = {{ACM}},
	series       = {{{CCS}} '11},
	pages        = {29--40},
	doi          = {10.1145/2046707.2046713},
	isbn         = {978-1-4503-0948-6},
	url          = {http://doi.acm.org/10.1145/2046707.2046713},
	urldate      = {2013-01-30},
	date         = 2011
}

@inproceedings {SHARD,
author = {Muhammad Abubakar and Adil Ahmad and Pedro Fonseca and Dongyan Xu},
title = {{SHARD}: {Fine-Grained} Kernel Specialization with {Context-Aware} Hardening},
booktitle = {30th USENIX Security Symposium (USENIX Security 21)},
year = {2021},
isbn = {978-1-939133-24-3},
pages = {2435--2452},
url = {https://www.usenix.org/conference/usenixsecurity21/presentation/abubakar},
publisher = {USENIX Association},
month = aug
}

@inproceedings{BASTION,
author = {Jelesnianski, Christopher and Ismail, Mohannad and Jang, Yeongjin and Williams, Dan and Min, Changwoo},
title = {Protect the System Call, Protect (Most of) the World with BASTION},
year = {2023},
isbn = {9781450399180},
publisher = {Association for Computing Machinery},
address = {New York, NY, USA},
url = {https://doi.org/10.1145/3582016.3582066},
doi = {10.1145/3582016.3582066},
booktitle = {Proceedings of the 28th ACM International Conference on Architectural Support for Programming Languages and Operating Systems, Volume 3},
pages = {528–541},
numpages = {14},
location = {Vancouver, BC, Canada},
series = {ASPLOS 2023}
}

@article{nacl3,
author = {Ansel, Jason and Marchenko, Petr and Erlingsson, \'{U}lfar and Taylor, Elijah and Chen, Brad and Schuff, Derek L. and Sehr, David and Biffle, Cliff L. and Yee, Bennet},
title = {Language-Independent Sandboxing of Just-in-Time Compilation and Self-Modifying Code},
year = {2011},
issue_date = {June 2011},
publisher = {Association for Computing Machinery},
address = {New York, NY, USA},
volume = {46},
number = {6},
issn = {0362-1340},
url = {https://doi.org/10.1145/1993316.1993540},
doi = {10.1145/1993316.1993540},
journal = {SIGPLAN Not.},
month = {jun},
pages = {355–366},
numpages = {12}
}

@inproceedings{NoNeedtoHide,
author = {Koning, Koen and Chen, Xi and Bos, Herbert and Giuffrida, Cristiano and Athanasopoulos, Elias},
title = {No Need to Hide: Protecting Safe Regions on Commodity Hardware},
year = {2017},
isbn = {9781450349383},
publisher = {Association for Computing Machinery},
address = {New York, NY, USA},
url = {https://doi.org/10.1145/3064176.3064217},
doi = {10.1145/3064176.3064217},
booktitle = {Proceedings of the Twelfth European Conference on Computer Systems},
pages = {437–452},
numpages = {16},
location = {Belgrade, Serbia},
series = {EuroSys '17}
}

@article{Kolosick22,
    author = {Kolosick, Matthew and Narayan, Shravan and Johnson, Evan and Watt, Conrad and LeMay, Michael and Garg, Deepak and Jhala, Ranjit and Stefan, Deian},
    title = {Isolation without Taxation: Near-Zero-Cost Transitions for WebAssembly and SFI},
    year = {2022},
    issue_date = {January 2022},
    publisher = {Association for Computing Machinery},
    address = {New York, NY, USA},
    volume = {6},
    number = {POPL},
    url = {https://doi.org/10.1145/3498688},
    doi = {10.1145/3498688},
    journal = {Proc. ACM Program. Lang.},
    month = {jan},
    articleno = {27},
    numpages = {30}
}

@INPROCEEDINGS{nacl1,
  author={Yee, Bennet and Sehr, David and Dardyk, Gregory and Chen, J. Bradley and Muth, Robert and Ormandy, Tavis and Okasaka, Shiki and Narula, Neha and Fullagar, Nicholas},
  booktitle={2009 30th IEEE Symposium on Security and Privacy}, 
  title={Native Client: A Sandbox for Portable, Untrusted x86 Native Code}, 
  year={2009},
  volume={},
  number={},
  pages={79-93},
  doi={10.1109/SP.2009.25}
}

@inproceedings {nacl2,
author = {David Sehr and Robert Muth and Cliff Biffle and Victor Khimenko and Egor Pasko and Karl Schimpf and Bennet Yee and Brad Chen},
title = {Adapting Software Fault Isolation to Contemporary {CPU} Architectures},
booktitle = {19th USENIX Security Symposium (USENIX Security 10)},
year = {2010},
address = {Washington, DC},
url = {https://www.usenix.org/conference/usenixsecurity10/adapting-software-fault-isolation-contemporary-cpu-architectures},
publisher = {USENIX Association},
month = aug
}

@online{apachebench,
	title        = {Ab - {{Apache HTTP}} Server Benchmarking Tool v2.3},
	url          = {https://httpd.apache.org/docs/2.4/en/programs/ab.html}
}

@online{nginx,
	title        = {{{NGINX}} v1.18.0},
	url          = {http://nginx.org/},
	organization = {{nginx}}
}

@misc{redis,
    author = {Redis Ltd.},
    title = {Redis},
    howpublished = {\url{https://redis.io/}},
    month = {Jan},
    year = {2024},
    note = {(Accessed on 01/16/2024)}
}

\end{document}